\documentclass[twoside,twocolumn,english,aps,prb,showpacs]{revtex4}

\usepackage[latin1]{inputenc}
\usepackage{graphicx}
\usepackage{amssymb}

\makeatletter

\usepackage{babel}
\makeatother
\begin{document}

\newcommand{\pd}{\hat{\psi}^{\dagger}}

\newcommand{\ps}{\hat{\psi}}
 
\newcommand{\ph}{\hat{\phi}}

\newcommand{\kt}[1]{\left|#1\right\rangle }

\title{A many-fermion generalization of the Caldeira-Leggett model}

\author{Florian Marquardt}

\affiliation{Department of Physics, Yale University, New Haven, CT 06520, USA}

\email{Florian.Marquardt@yale.edu}

\author{D.S. Golubev}

\affiliation{Institut für Theoretische Festkörperphysik, Universität Karlsruhe,
76128 Karlsruhe, Germany }

\affiliation{I. E. Tamm Department of Theoretical Physics, P. N. Lebedev Physics
Institute, 119991 Moscow, Russia }

\date{15.11.2004}

\begin{abstract}
We analyze a model system of fermions in a harmonic oscillator potential
under the influence of a dissipative environment: The fermions are
subject to a fluctuating force deriving from a bath of harmonic oscillators.
This represents an extension of the well-known Caldeira-Leggett model
to the case of many fermions. Using the method of bosonization, we
calculate one- and two-particle Green's functions of the fermions.
We discuss the relaxation of a single extra particle added above the
Fermi sea, considering also dephasing of a particle added in a coherent
superposition of states. The consequences of the separation of center-of-mass
and relative motion, the Pauli principle, and the bath-induced effective
interaction are discussed. Finally, we extend our analysis to a more
generic coupling between system and bath, that results in complete
thermalization of the system.
\end{abstract}

\pacs{03.65.Yz, 05.30.Fk, 71.10.Pm}

\maketitle

\section{Introduction}

The study of quantum systems subject to dissipative environments is
a topic which is both of fundamental importance in quantum mechanics
and relevant for many applications requiring quantum-coherent dynamics.
Friction, energy relaxation, thermalization, destruction of quantum
interference effects (decoherence) and the irreversibility of the
measurement process are all examples of features that arise due to
the system-environment coupling. The most important exactly solvable
model in the theory of quantum-dissipative systems\cite{weiss} is
the Caldeira-Leggett (CL) model\cite{callegg,clwavepackets}. It is
the simplest quantum-mechanical model describing friction and fluctuations,
and it has been used to analyze the quantum analogue of Brownian motion.
The model consists of a single particle whose coordinate is coupled
bilinearly to the coordinates of a bath of harmonic oscillators. Thus,
it can be solved exactly as long as the particle either moves freely
(apart from the system-bath coupling) or inside a harmonic oscillator
potential. 

However, in several important applications of these concepts we are
not dealing with a single particle but rather with a many-particle
problem from the outset. This applies in particular to solid state
physics, where the dephasing of electronic motion often determines
the temperature-dependence of quantum-mechanical interference effects
such as Aharonov-Bohm interference, universal conductance fluctuations,
or weak localization. Usually, the fermionic nature of the electrons
makes it difficult to apply the insights gained from single-particle
calculations, since the Pauli principle may play an important role
in relaxation processes. %
\begin{figure}
\begin{center}\includegraphics[%
  width=1.0\columnwidth]{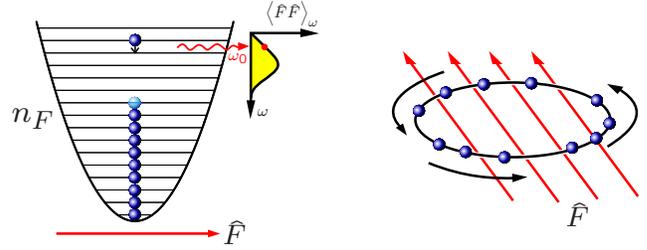}\end{center}

\caption{\label{DFSfigharmon}Left: Coupling the fermions in the oscillator
to the fluctuating quantum force $\hat{F}$ leads both to relaxation
of excited fermions (decay rates depending on the bath spectrum $\left\langle \hat{F}\hat{F}\right\rangle _{\omega}$),
as well as some smearing of the level occupation even at $T=0$. Right:
The bosonized version is equivalent to chiral fermions moving on a
ring, subject to a transverse fluctuating force.}
\end{figure}

In this paper, we present our results for a natural extension of the
CL model to a case with many fermions. We consider a number of non-interacting
fermions moving inside a harmonic oscillator potential, subject to
a fluctuating quantum force deriving from an oscillator bath (Fig.
\ref{DFSfigharmon}). In the single-particle case, this is just the
CL model, describing a quantum damped harmonic oscillator. Thus, we
are studying a {}``Fermi sea in a damped harmonic oscillator''.
Our main motivation is to analyze an exactly solvable model problem
featuring dephasing and relaxation in presence of the Pauli principle.
Nevertheless, the model may also become helpful for the study of cold
fermionic atom clouds in quasi one-dimensional harmonic atom traps,
where fluctuations of the external trapping potential might be described
by a homogeneous fluctuating force acting on the cloud. Alternatively,
it could be an approximate description of a parabolic (quasi one-dimensional)
quantum dot subject to the Nyquist noise of a fluctuating electric
field from nearby gates. 

There are only a few previous theoretical works studying dephasing
and relaxation of fermions in the context of quantum-dissipative system
(excluding Boltzmann-type kinetic equations or simple perturbation
theory). To the best of our knowledge, a many-fermion generalization
of the CL model was first suggested in Ref. \onlinecite{indepFermions},
where free fermions coupled to \emph{independent} oscillator baths
were studied. As this breaks the indistinguishability of the fermions,
it is not clear to which extent this model may be realized physically,
and the influence of Pauli-blocking on relaxation processes could
not be studied. A discussion of Luttinger liquids coupled to dissipative
environments has been provided in Ref. \onlinecite{OpenLuttLiquids},
where the emphasis was on general features rather than the actual
evaluation of Green's functions. Some aspects of our model are similar
to the features found in a study\cite{wonneberger} of interacting
fermions in a parabolic trap without coupling to a bath. Another extension
of the CL model to many fermions\cite{ABring} involved particles
on a ring subject to a quantum force independent of the position on
the circumference. However, this did not reveal any influence of the
Pauli principle, as such a coupling does not lead to transitions between
different momentum states (Pauli blocking becomes relevant only for
tunneling from external Fermi systems). Coupling a heat bath to indistinguishable
fermions has also been employed as a tool for quantum molecular dynamics\cite{schnack}.
For a recent detailed perturbative study of Nyquist noise leading
to decoherence of electrons in quantum dots and wires, for realistic
gate geometries, we refer the reader to Refs. \onlinecite{guinea}.
The Feynman-Vernon influence functional, widely used in the analysis
of a single dissipative particle, cannot be applied directly to the
many-fermion situation. In the theory of weak-localization, there
have been phenomenological prescriptions to incorporate the effects
of the Pauli principle in an influence functional approach\cite{AAKetc}.
Recently, a formally exact generalization to the many-fermion case\cite{FV-Fermi,janFV}
has been derived, although the evaluation of the resulting path-integral
remains difficult. 

The work is organized as follows: After introducing the Hamiltonian
in Sec. \ref{sec:The-model}, we apply a certain approximation in
order to rewrite and solve the Hamiltonian using the method of bosonization,
in Sec. \ref{bosonAndDiagon}. This forms the basis for our evaluation
of the fermions' single-particle Green's functions, described in Secs.
\ref{DFSgreensfctT0} and \ref{sec:Discussion-of-the}, leading to
the exact general expression in Eqs. (\ref{DFSg0E}), (\ref{DFSexponent}).
In Section \ref{sec:Finite-particle-number:}, we analyze in more
detail our approximation of 'large' particle numbers needed for bosonization.
In Section \ref{DFStwoparticlemaintext}, we evaluate the two-particle
Green's function (exact result in Eqs. (\ref{DFSa0etilde})-(\ref{DFSetilde})),
which enables us to discuss the decay of populations after adding
an extra single particle above the Fermi sea, as well as dephasing
of a coherent superposition of states. Finally, we extend our model
to a more generic type of coupling between fermions and bath (Sec.
\ref{DFSrelaxarbitraryenergytransfer}). Technical details are relegated
to the appendices.

Some of these results have already been presented in a brief version\cite{FlorianDima}.

\section{The model}

\label{sec:The-model}We consider a system of $N$ identical fermions
(non-interacting and spinless) confined in a one-dimensional harmonic
oscillator potential. They are subject to a fluctuating force $\hat{F}$
that by necessity acts on each fermion with equal strength, i.e. it
couples to the center-of-mass coordinate of the system of fermions
(see Fig. \ref{DFSfigharmon}), like $\propto\hat{F}\sum_{j}\hat{x}_{j}$.
This force derives from a bath of oscillators. In second quantiziation,
the Hamiltonian reads:

\begin{eqnarray}
\hat{H}=\omega_{0}\sum_{n=0}^{\infty}n\hat{c}_{n}^{\dagger}\hat{c}_{n}+\hat{H}_{B}+\nonumber \\
\frac{\hat{F}}{\sqrt{2m\omega_{0}}}\sum_{n=0}^{\infty}\sqrt{n+1}(\hat{c}_{n+1}^{\dagger}\hat{c}_{n}+h.c.)\label{DFSham}\end{eqnarray}
The $\hat{c}_{n}$ are fermion annihilation operators, and the oscillation
frequency of a fermion of mass $m$ in the parabolic potential is
$\omega_{0}$ (we have set $\hbar=1$). Note that $\omega_{0}$ already
contains a counterterm that depends on the coupling to the bath, see
Eq. (\ref{DFSfreqbare}) below. The bath Hamiltonian is given by:

\begin{equation}
\hat{H}_{B}=\sum_{j=1}^{N_{B}}\frac{\hat{P}_{j}^{2}}{2}+\frac{m\Omega_{j}^{2}}{2}\hat{Q}_{j}^{2}\,.\end{equation}
Here the bath oscillator masses have been chosen to be equal to that
of the fermions, without any loss of generality, in order to streamline
a few expressions derived in the next section. The force $\hat{F}$
is given as a sum over the bath normal coordinates $\hat{Q}_{j}$
(with a prefactor $g$ of dimensions energy over length squared):

\begin{equation}
\hat{F}=\frac{g}{\sqrt{N_{B}}}\sum_{j=1}^{N_{B}}\hat{Q}_{j}\,.\end{equation}
Its spectrum 

\begin{equation}
\left\langle \hat{F}\hat{F}\right\rangle _{\omega}=\frac{1}{2\pi}\int_{-\infty}^{+\infty}\left\langle \hat{F}(t)\hat{F}\right\rangle e^{+i\omega t}dt\,\end{equation}
is still arbitrary and depends on the distribution of bath oscillator
frequencies. At $T=0$, it is given by:

\begin{equation}
\left\langle \hat{F}\hat{F}\right\rangle _{\omega}^{T=0}=\frac{g^{2}}{N_{B}}\sum_{j}\frac{1}{2m\Omega_{j}}\delta(\omega-\Omega_{j})\,.\label{DFSfft0}\end{equation}
The special case of an Ohmic bath, as it is used in the theory of
Quantum Brownian motion \cite{callegg}, is defined by

\begin{equation}
\left\langle \hat{F}\hat{F}\right\rangle _{\omega}^{T=0}=\frac{\eta}{\pi}\omega\theta(\omega_{c}-\omega)\theta(\omega)\,,\label{DFSohmicbath}\end{equation}
where $\eta=m\gamma$ is the coefficient entering the friction force
$-\eta v$ acting on a single particle of mass $m$, with $\gamma$
the corresponding damping rate. 

The Hamiltonian (\ref{DFSham}) can be derived from the following
form, where the fermions are treated without second quantization and
the translational invariance of the coupling between fermions and
bath particles is apparent:

\begin{eqnarray}
\sum_{l=1}^{N}\frac{\hat{p}_{l}^{2}}{2m}+\frac{m\omega_{00}^{2}}{2}\hat{x}_{l}^{2}+\nonumber \\
\frac{1}{N}\sum_{l=1}^{N}\sum_{j=1}^{N_{B}}g_{j}(x_{l}-\hat{\tilde{Q}}_{j})^{2}+\sum_{j=1}^{N_{B}}\frac{\hat{P}_{j}^{2}}{2m}\label{DFSx2}\end{eqnarray}
Here the couplings $g_{j}$ are given in terms of the parameter $g$
as $g_{j}=g^{2}N^{2}/(2m\Omega_{j}^{2}N_{B})$, and the rescaled bath
coordinates are $\hat{\tilde{Q}}_{j}=\hat{Q}_{j}(m\Omega_{j}^{2}\sqrt{N_{B}}/(gN))$.
Note that the frequency $\omega_{0}$ introduced above already contains
the counterterm that arises from the $x_{l}^{2}$-terms in (\ref{DFSx2})
and which is essential to prevent the effective oscillator potential
from becoming unstable for larger coupling strengths $g$. In terms
of the bare oscillator frequency $\omega_{00}$, it is given by:

\begin{equation}
\omega_{0}^{2}=\omega_{00}^{2}+2\frac{N}{m}\int_{0}^{\infty}d\omega\,\frac{\left\langle \hat{F}\hat{F}\right\rangle _{\omega}^{T=0}}{\omega}\,.\label{DFSfreqbare}\end{equation}

\section{Solution by bosonization and diagonalisation}

\label{bosonAndDiagon}Since the fluctuating force acts only on the
center-of-mass (c.m.) coordinate of the particles and, in the case
of the harmonic oscillator potential, the c.m. motion is independent
of the relative motion, the model defined above is, in principle,
exactly solvable in a straightforward manner. The solution can be
carried out by finding the classical eigenfrequencies and eigenvectors
of the total system of $N+N_{B}$ coupled oscillators, setting up
the quantum-mechanical wave functions in the total Hilbert space and
performing the antisymmetrization with respect to the fermion coordinates.
Afterwards, any desired observable (reduced density matrix, occupation
numbers, fermion Green's functions etc.) may be calculated in principle.
However, due to the antisymmetrization and the appearance of the oscillator
eigenfunctions in the intermediate steps of the calculation, this
procedure gets extremely cumbersome, so we prefer a different route
which is approximately valid for large fermion numbers $N$. In Section
\ref{sec:Finite-particle-number:}, we will analyze some aspects of
the small $N$ case and compare with the bosonization results.

We assume the number $N$ of fermions in the oscillator potential
to be so large that the lowest-lying oscillator states are always
occupied, for any given many-particle state that becomes relevant
in the calculation (at the given interaction strength and temperatures).
In other words, the excitations in the fermion system, induced by
the bath (and temperature), are confined to the region near the Fermi
surface. Then we may employ the method of bosonization where the energy
of the fermions is rewritten as a sum over boson modes (i.e. sound
waves in the fermion system). This is possible since the energies
of the oscillator levels increase linearly with quantum number, just
as the kinetic energy of electrons in the Luttinger model of interacting
electrons in one dimension. The same procedure has been applied to
describe interacting fermionic atoms in a parabolic one-dimensional
trap, for certain exactly solvable model interactions\cite{wonneberger}.
For recent pedagogical reviews on Luttinger liquids and bosonization,
see Ref. \onlinecite{DelftSchoeller}. We note that this equal spacing
of energy levels is approximately present for any potential at high
excitation energies, where a semiclassical description of the single-particle
dynamics becomes valid. However, the applicability of the following
description to such a situation also depends on the structure of matrix
elements of the fluctuating potential in the single-particle eigenbasis.
For the harmonic oscillator considered here, only adjacent levels
are connected by the position operator $\hat{x}$ (describing the
potential of the homogeneous force).

We introduce (approximate) boson operators, which destroy particle-hole
excitations ($q\geq1$):

\begin{equation}
\hat{b}_{q}=\frac{1}{\sqrt{q}}\sum_{n=0}^{\infty}\hat{c}_{n}^{\dagger}\hat{c}_{n+q}\,.\end{equation}
One may check that they fulfill the usual boson commutation relations,
up to terms involving levels near $n=0$ that vanish when acting on
the many-particle states occuring under our assumptions. Alternatively,
one may make these relations hold exactly by redefining the original
model to incorporate an infinite number of artificial single-particle
levels of negative energy, as is done in the Luttinger model. 

Then the central result of bosonization may be applied, i.e. the fermion
energy (bilinear in $\hat{c}_{n}^{(\dagger)}$) may be written as
a bilinear expression in $\hat{b}_{q}^{(\dagger)}$. This is possible
only due to the linear dependence of energies on the quantum number
$n$:

\begin{equation}
\omega_{0}\sum_{n=0}^{\infty}n\hat{c}_{n}^{\dagger}\hat{c}_{n}=\omega_{0}\sum_{q=1}^{\infty}q\hat{b}_{q}^{\dagger}\hat{b}_{q}+E_{\hat{N}}\,.\end{equation}
Here $E_{\hat{N}}=\omega_{0}\hat{N}(\hat{N}-1)/2$ is the total energy
of the $N$-fermion noninteracting ground state. We keep $\hat{N}$
as an operator at this point since we will be interested in calculating
Green's functions where the particle number changes. 

Under the same assumption of large $N\gg1$ we get:

\begin{equation}
\sum_{n=0}^{\infty}\sqrt{n+1}(\hat{c}_{n+1}^{\dagger}\hat{c}_{n}+h.c.)\approx\sqrt{N}(\hat{b}_{1}+\hat{b}_{1}^{\dagger})\,.\end{equation}
Again, this has to be understood as an approximate operator identity
which is valid when applied to the many-particle states we are interested
in, where $n\approx N$. We approximate $N$ to be a number instead
of an operator in this formula. This expression shows that the fluctuating
force only couples to the lowest boson mode (with $q=1$), corresponding
to the c.m. motion. Therefore, the Hamiltonian now has become (within
the approximations described above):

\begin{equation}
\hat{H}\approx\omega_{0}\sum_{q=1}^{\infty}q\hat{b}_{q}^{\dagger}\hat{b}_{q}+\sqrt{\frac{N}{2m\omega_{0}}}\hat{F}(\hat{b}_{1}+\hat{b}_{1}^{\dagger})+\hat{H}_{B}+E_{\hat{N}}\,.\label{DFShapprox}\end{equation}
This Hamiltonian constitutes the starting point for our subsequent
analysis. It can be solved by diagonalization of the (classical) problem
of the boson oscillator $q=1$ coupled to the bath oscillators (see
Appendix \ref{DFSappendixDiagon} and below, also Ref. \onlinecite{hakim}).
Note that coupling of a Luttinger liquid to a linear bath has been
considered in Ref. \onlinecite{OpenLuttLiquids}, although the physics
discussed there (as well as the calculation) is quite distinct from
our model.

At the end we are interested in quantities relating to the fermions
themselves, e.g. the occupation numbers $\left\langle \hat{c}_{n}^{\dagger}\hat{c}_{n}\right\rangle $
or the Green's functions, like $\left\langle \hat{c}_{n}(t)\hat{c}_{n}^{\dagger}(0)\right\rangle $.
This means we have to go back from the boson operators $\hat{b}_{q}$
to the fermion operators, by employing the relations which are also
used in the Luttinger liquid. In order to do that, we have to introduce
auxiliary fermion operators $\hat{\psi}(x)$:

\begin{eqnarray}
\hat{\psi}(x)=\frac{1}{\sqrt{2\pi}}\sum_{n}e^{inx}\hat{c}_{n}\\
\hat{c}_{n}=\frac{1}{\sqrt{2\pi}}\int_{0}^{2\pi}e^{-inx}\hat{\psi}(x)\, dx\label{DFScpsi}\end{eqnarray}
Note that the $\hat{\psi}(x)$ are \emph{not} directly related to
the fermion operators of the particles in the oscillator (which would
involve the oscillator eigenfunctions). We have effectively mapped
our problem to a chiral Luttinger liquid on a ring with a coupling
$\propto\hat{F}\cos(x)$ ($x\in[0,2\pi[$), see Fig. \ref{DFSfigharmon}
(right). The following results also describe relaxation of momentum
states in that model. The $\hat{\psi}$ operators are useful because
they fulfill

\begin{equation}
\left[\hat{b}_{q},\,\hat{\psi}(x)\right]=-\frac{1}{\sqrt{q}}e^{-iqx}\hat{\psi}(x)\,.\label{DFScommutator}\end{equation}
This means the application of $\hat{\psi}(x)$ on the noninteracting
$N$-particle ground state creates an $N-1$-particle state that is
a coherent state with respect to the boson modes, i.e. an eigenstate
of $\hat{b}_{q}$ for every $q$. As a consequence, the fermion operators
$\hat{\psi}(x)$ may be expressed as \cite{DelftSchoeller}:

\begin{equation}
\hat{\psi}(x)=\hat{K}\lambda(x)e^{i\hat{\varphi}^{\dagger}(x)}e^{i\hat{\varphi}(x)}=\hat{K}\lambda e^{i\hat{\phi}}r\,,\label{DFSpsibos}\end{equation}
with

\begin{eqnarray}
\hat{\varphi}(x)=-i\sum_{q=1}^{\infty}\frac{1}{\sqrt{q}}e^{iqx}\hat{b}_{q}\label{DFSphidef}\\
\hat{\phi}=\hat{\varphi}+\hat{\varphi}^{\dagger}\label{DFSPHIdef}\\
r\equiv e^{-[\hat{\varphi}^{\dagger},\hat{\varphi}]/2}\,.\end{eqnarray}
The exponential $\exp(i\hat{\varphi}^{\dagger}(x))$ in (\ref{DFSpsibos})
may be recognized as creating a coherent state with the eigenvalues
of $\hat{b}_{q}$ prescribed by Eq. (\ref{DFScommutator}). The other
terms are necessary to give the correct normalization and phase-factor,
and to deal with states other than the $N$-particle ground state.

The second equality in (\ref{DFSpsibos}) follows from the Baker-Hausdorff
identity. The {}``Klein factor'' $\hat{K}$ is defined to commute
with the boson operators $\hat{b}_{q}^{(\dagger)}$ and to produce
the noninteracting $(N-1)$-particle ground state out of the noninteracting
$N$-particle ground state. Its time-evolution follows from $[\hat{K},\hat{H}]=[\hat{K},E_{\hat{N}}]=\hat{K}\omega_{0}(\hat{N}-1)$
as $\hat{K}(t)=\hat{K}\exp(-i\omega_{0}(\hat{N}-1)t)$. The factor
$\lambda(x)$ is given by $\exp(i(\hat{N}-1)x)/\sqrt{2\pi}$. Actually,
$[\hat{\varphi}^{\dagger},\hat{\varphi}]=-\sum_{q=1}^{\infty}1/q$
diverges, so we would have to introduce an artificial cutoff $e^{-aq}$
($a\rightarrow0$) into the sum. However, this drops out in the end
result, because $r$ from Eq. (\ref{DFSpsibos}) is canceled by the
contributions from the equal-time $\ph$-correlators in the exponent
of Eq. (\ref{DFSpsigr}). 

Using the relation (\ref{DFScpsi}), we have, for example:

\begin{equation}
\left\langle \hat{c}_{n}^{\dagger}(t)\hat{c}_{n}\right\rangle =\frac{1}{2\pi}\int_{0}^{2\pi}e^{in(x'-x)}\left\langle \hat{\psi}^{\dagger}(x',t)\hat{\psi}(x,0)\right\rangle \, dx\, dx'\,.\label{DFSctc}\end{equation}
The Green's function involving $\hat{\psi}$ may be expressed directly
in terms of the correlator of the boson operator $\hat{\phi}$, using
Eq. (\ref{DFSpsibos}):

\begin{eqnarray}
 &  & \left\langle \pd(x',t)\ps(x,0)\right\rangle =\frac{1}{2\pi}e^{in_{F}((x-x')+\omega_{0}t)}\times\nonumber \\
 &  & \left\langle e^{-i\hat{\phi}(x',t)}e^{i\hat{\phi}(x,0)}\right\rangle r^{2}\,.\label{DFSpsigr}\end{eqnarray}
We have used the abbreviation $n_{F}=N-1$ for the quantum number
of the highest occupied state (in the noninteracting system). The
expectation value on the right-hand side is evaluated using the Baker-Haussdorff
identity and the well-known Gaussian property of the bosonic variables
(which is unchanged by the coupling to the linear bath). This yields:

\begin{equation}
\exp\left[-\frac{1}{2}\left(\left\langle \hat{\phi}(x',t)^{2}\right\rangle +\left\langle \ph(x,0)^{2}\right\rangle \right)+\left\langle \ph(x',t)\ph(x,0)\right\rangle \right]\label{DFSexpphi}\end{equation}
These results permit us to calculate the hole propagator $\left\langle \hat{c}_{n}^{\dagger}(t)\hat{c}_{n}\right\rangle $,
from which we obtain the equilibrium density matrix by setting $t=0$.
Note that we have particle-hole symmetry in our problem, such that
the particle propagator gives no additional information. Writing down
the expressions analogous to (\ref{DFSctc}) and (\ref{DFSpsigr}),
and using the properties of $\lambda(x)$ and $\hat{K}(t)$ defined
above, we find:

\begin{eqnarray}
\left\langle \hat{c}_{n_{F}+\delta n}(t)\hat{c}_{n_{F}+\delta n}^{\dagger}\right\rangle =\nonumber \\
\left\langle \hat{c}_{n_{F}+1-\delta n}^{\dagger}(t)\hat{c}_{n_{F}+1-\delta n}\right\rangle e^{-i\omega_{0}(2n_{F}+1)t}\,.\end{eqnarray}
 It remains to calculate the correlator of $\ph$ in the interacting
equilibrium: This is where the coupling to the bath enters, since
the original $q=1$ mode will get mixed with the bath modes. All the
other boson modes are unaffected. The different boson modes remain
independent. We obtain (using Eqs. (\ref{DFSphidef}) and (\ref{DFSPHIdef})):

\begin{eqnarray}
 &  & \left\langle \ph(x',t)\ph(x,0)\right\rangle =\nonumber \\
 &  & -\sum_{q=1}^{\infty}\frac{1}{q}\left\langle (e^{iqx'}\hat{b}_{q}(t)-h.c.)(e^{iqx}\hat{b}_{q}-h.c.)\right\rangle \,.\label{DFSphicorr}\end{eqnarray}
The expectation values for $q>1$ are the original ones, for example:

\begin{equation}
\left\langle \hat{b}_{q}(t)\hat{b}_{q}^{\dagger}\right\rangle =e^{-i\omega_{0}qt}(n(\omega_{0}q)+1)\,,\end{equation}
with $n(\epsilon)=(e^{\beta\epsilon}-1)^{-1}$ the Bose distribution
function.

In order to obtain the time-evolution and equilibrium expectation
values relating to the $q=1$ mode (c.m. mode), we have to diagonalize
a quadratic Hamiltonian describing the coupling between this mode
and the bath oscillators, which is described in Appendix \ref{DFSappendixDiagon}.

\section{Evaluation of the Green's function for $T=0$}

\label{DFSgreensfctT0}At $T=0$, some simplifications apply to the
evaluation of the $\hat{\phi}$-correlator and, consequently, the
hole propagator $\left\langle \hat{c}_{n}^{\dagger}(t)\hat{c}_{n}\right\rangle $.

In Appendix \ref{DFSappendixDiagon}, we express the desired c.m.
correlator in terms of the uncoupled normal mode operators. The result
can be written in terms of the eigenvalues $\tilde{\Omega_{j}}$ and
corresponding eigenvectors $C_{\cdot j}$:

\begin{eqnarray}
\left\langle \hat{b}_{1}(t)\hat{b}_{1}\right\rangle =\frac{1}{4}\sum_{j=0}^{N_{B}}C_{0j}^{2}\left(\frac{\omega_{0}}{\tilde{\Omega}_{j}}-\frac{\tilde{\Omega}_{j}}{\omega_{0}}\right)e^{-i\tilde{\Omega}_{j}t}\,.\label{DFSgrfct}\end{eqnarray}
By defining the spectral weight $W(\omega)$ of the c.m. mode (entry
$0$ in this notation, see App. \ref{DFSappendixDiagon}) in the new
eigenbasis, 

\begin{equation}
W(\omega)=\sum_{j=0}^{N_{B}}C_{0j}^{2}\delta(\omega-\tilde{\Omega}_{j})\,,\label{DFSWdef}\end{equation}
which is normalized to $1$, we can rewrite (\ref{DFSgrfct}) as:

\begin{equation}
\left\langle \hat{b}_{1}(t)\hat{b}_{1}\right\rangle =\frac{1}{4}\int_{0}^{\infty}W(\omega)\,(\frac{\omega_{0}}{\omega}-\frac{\omega}{\omega_{0}})e^{-i\omega t}\, d\omega\equiv\alpha(t)\,.\label{DFSalpha}\end{equation}
In a completely analogous fashion, we find:

\begin{eqnarray}
 &  & \left\langle \hat{b}_{1}(t)\hat{b}_{1}^{\dagger}\right\rangle \equiv\beta_{+}(t)=\label{DFSbetaplus}\\
 &  & \frac{1}{4}\int_{0}^{\infty}W(\omega)\,\left(\sqrt{\frac{\omega_{0}}{\omega}}+\sqrt{\frac{\omega}{\omega_{0}}}\right)^{2}e^{-i\omega t}\, d\omega\nonumber \\
 &  & \left\langle \hat{b}_{1}^{\dagger}(t)\hat{b}_{1}\right\rangle \equiv\beta_{-}(t)=\label{DFSbetaminus}\\
 &  & \frac{1}{4}\int_{0}^{\infty}W(\omega)\,\left(\sqrt{\frac{\omega_{0}}{\omega}}-\sqrt{\frac{\omega}{\omega_{0}}}\right)^{2}e^{-i\omega t}\, d\omega,\nonumber \end{eqnarray}
and $\left\langle \hat{b}_{1}^{\dagger}(t)\hat{b}_{1}^{\dagger}\right\rangle =\left\langle \hat{b}_{1}(t)\hat{b}_{1}\right\rangle =\alpha(t)$. 

The function $W(\omega)$ is derived in Appendix \ref{DFSappendixDiagon},
in terms of the bath spectrum. The result is

\begin{equation}
W(\omega)=\frac{\omega}{\pi}\theta(\omega)\frac{\Gamma(\omega^{2})}{(\omega^{2}-\omega_{0}^{2}-\Delta(\omega^{2}))^{2}+(\Gamma(\omega^{2})/2)^{2}}\,,\label{DFSWresult}\end{equation}
with:

\begin{eqnarray}
\Gamma(\epsilon) & = & 2\pi\frac{N}{m}\left\langle \hat{F}\hat{F}\right\rangle _{\sqrt{\epsilon}}^{T=0}\theta(\epsilon)\label{DFSappGammaDef}\\
\Delta(\epsilon) & = & \frac{1}{2\pi}\int\frac{\Gamma(\nu)}{\epsilon-\nu}d\nu\,.\end{eqnarray}
(a principal value integral is implied in the last line).

For weak coupling, $W(\omega)$ is a Lorentz peak centered around
$\omega_{0}$. That is why $\alpha(t)$ and $\beta_{-}(t)$ are small,
since the spectrum $W(\omega)$ is multiplied by a function that vanishes
at the resonance near $\omega_{0}$. Both $\alpha$ and $\beta_{-}$
vanish exactly at zero coupling. In contrast, $\beta_{+}(t)$ describes
damped oscillations around $\omega_{0}$ (see below for a discussion
of the damping rate), starting from $\beta_{+}(0)\approx1$. 

Evaluation of the c.m. mode contribution to the correlator of $\hat{\phi}$
(given in Eq. (\ref{DFSphicorr})) then leads to the following expression:

\begin{eqnarray}
\left\langle (e^{ix'}\hat{b}_{1}(t)-h.c.)(e^{ix}\hat{b}_{1}-h.c.)\right\rangle =\nonumber \\
(e^{i(x'+x)}+c.c.)\alpha(t)-e^{i(x'-x)}\beta_{+}(t)-e^{i(x-x')}\beta_{-}(t)\,.\label{DFSabb}\end{eqnarray}
Now we are prepared to evaluate the Green's functions of the fermions
in the damped oscillator. It is convenient to introduce the abbreviations:

\begin{equation}
X\equiv e^{ix},\, X'\equiv e^{ix'}\,.\end{equation}
Then Eq. (\ref{DFSabb}) leads to:

\begin{eqnarray}
\left\langle \ph(x',t)\ph(x,0)\right\rangle =\left\langle \ph(x',t)\ph(x,0)\right\rangle _{(0)}+\nonumber \\
\frac{X'}{X}(\beta_{+}(t)-e^{-i\omega_{0}t})+\frac{X}{X'}\beta_{-}(t)-\nonumber \\
\alpha(t)(XX'+(XX')^{-1})\,,\label{DFSphiXX}\end{eqnarray}
where the subscript $(0)$ refers to the case without coupling to
a bath:

\begin{equation}
\left\langle \ph(x',t)\ph(x,0)\right\rangle _{(0)}=\sum_{q=1}^{\infty}\frac{(e^{-i\omega_{0}t}X'/X)^{q}}{q}\,.\label{DFSphi0qsum}\end{equation}
We insert this expression for the $\ph$-correlator into the exponent
(\ref{DFSexpphi}) that appears in the $\hat{\psi}$ Green's function,
Eq. (\ref{DFSpsigr}), and perform the Fourier transform (\ref{DFSctc})
with respect to $x,x'$ to go back to the original oscillator fermion
operators $\hat{c}_{n}$. 

Then we find that the hole propagator $\left\langle \hat{c}_{n_{F}-\delta n'}^{\dagger}(t)\hat{c}_{n_{F}-\delta n}\right\rangle $
is given by the prefactor of $X'^{\delta n'}/X^{\delta n}$ in the
power series generated from 

\begin{eqnarray}
\sum_{k=0}^{\infty}(X'/X)^{k}e^{i\omega_{0}(n_{F}-k)t}e^{\delta E(X,X',t)},\label{DFSg0E}\end{eqnarray}
with\begin{eqnarray}
\delta E(X,X',t)\equiv1-\beta_{+}(0)-\beta_{-}(0)+\nonumber \\
\frac{\alpha(0)}{2}(X'^{2}+X^{2}+X'^{-2}+X^{-2})-\nonumber \\
(X'X+1/(X'X))\alpha(t)+\nonumber \\
(X'/X)(\beta_{+}(t)-e^{-i\omega_{0}t})+(X/X')\beta_{-}(t)\,.\label{DFSexponent}\end{eqnarray}
Equation (\ref{DFSg0E}) constitutes our central exact result for
the single-particle Green's function of the bosonized model. 

The exponent $\delta E$ describes the influence of the bath on the
Green's function. It follows by inserting (\ref{DFSphiXX}) into the
exponential Eq. (\ref{DFSexpphi}). The non-interacting value of the
exponent has been split off and is accounted for by the prefactor
in Eq. (\ref{DFSg0E}), which reproduces the correct result for the
non-interacting case, namely $\left\langle \hat{c}_{n_{F}-\delta n}^{\dagger}(t)\hat{c}_{n_{F}-\delta n}\right\rangle =e^{i(n_{F}-\delta n)\omega_{0}t}$
for $\delta n\geq0$.

The weak-coupling approximation consists in setting $\alpha(t)=\beta_{-}(t)=0$,
$\beta_{+}(0)=1$ and keeping only the (slowly decaying) $\beta_{+}(t)$,
which is approximated by $\beta_{+}(t)\approx e^{-i\omega'_{0}t-N\gamma t/2}$.
Here $\omega'_{0}$ is shifted with respect to the frequency $\omega_{0}$,
being equal to $\omega_{00}$ in the limit $\omega_{c}\gg\omega_{0}$
(see Eqs. (\ref{DFSohmbathomega}) and (\ref{DFSohmW}) for the special
case of the Ohmic bath). In this approximation, we obtain:

\begin{equation}
\left\langle \hat{c}_{n}^{\dagger}(t)\hat{c}_{n}\right\rangle \approx e^{i\omega_{0}nt}\sum_{m=0}^{n_{F}-n}\frac{\nu(t)^{m}}{m!}\,,\label{weakcouplGF}\end{equation}
 where \begin{equation}
\nu(t)\equiv\exp(-i(\omega_{0}^{'}-\omega_{0})t-N\gamma t/2)-1\,,\label{nu}\end{equation}
 with the decay rate $N\gamma$ evaluated for the Ohmic bath.

It is even possible to derive a closed integral expression for the
exact single-particle Green's function. We only state the result:

\begin{eqnarray}
\left\langle \hat{c}_{n'}^{\dagger}(t)\hat{c}_{n}\right\rangle  & = & \frac{1+(-1)^{n'-n}}{4\pi}e^{i(n+n')\omega_{0}t/2-2\beta_{-}(0)}\times\nonumber \\
 &  & \int_{-\pi}^{\pi}dx\,\frac{e^{i(n_{F}-(n'+n)/2)x}}{1+0^{+}-e^{-ix}}\times\nonumber \\
 &  & e^{[e^{i\omega_{0}t}\beta_{+}(t)-1]e^{-ix}+e^{-i\omega_{0}t}\beta_{-}(t)e^{ix}}\times\nonumber \\
 &  & I_{\frac{n'-n}{2}}[2\alpha(0)\cos(x-\omega_{0}t)-2\alpha(t)].\label{GreensFunctExt}\end{eqnarray}
Here $I_{m}$ is the modified Bessel function of the first kind. A
useful approximation becomes possible in the limit $n\ll n_{F}$,
i.e. for a high excitation energy, even at strong coupling:

\begin{eqnarray}
\left\langle \hat{c}_{n_{F}-\delta n}^{\dagger}(t)\hat{c}_{n_{F}-\delta n}\right\rangle \approx\nonumber \\
\exp(e^{i\omega_{0}t}\beta_{+}(t)-\beta_{+}(0)+e^{-i\omega_{0}t}\beta_{-}(t)-\beta_{-}(0))\times\nonumber \\
e^{i\omega_{0}(n_{F}-\delta n)t}I_{0}[2\alpha(0)\cos(\omega_{0}t)-2\alpha(t)]\,.\end{eqnarray}
Note the trivial dependence on $\delta n$ in this limit (a shift
in frequency). As $I_{0}(x=0)=1$, we can easily recover the weak-coupling
form from above.

\section{Discussion of equilibrium density matrix and Green's function}

\label{sec:Discussion-of-the}
\begin{figure}
\includegraphics[%
  width=1.0\columnwidth]{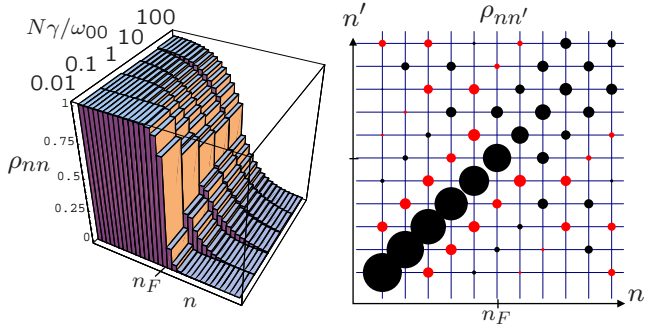}

\caption{\label{cap:Equilibrium-density-matrix:}Equilibrium density matrix.
Left: Populations $\rho_{nn}$ as a function of coupling strength.
Right: Full density matrix $\rho_{nn'}$ for fixed $N\gamma/\omega_{00}=10$,
magnitude indicated by radius, negative values indicated by different
color/gray level. Note: Off-diagonal elements have been enlarged by
a factor of $3$ for clarity. We have chosen $\omega_{c}/\omega_{00}=10^{4}$
in both plots. }
\end{figure}

Setting $t=0$ in Eq. (\ref{GreensFunctExt})
yields the equilibrium density matrix $\rho_{nn'}$, for which only
$\alpha(0),$ $\beta_{-}(0)$ and $\beta_{+}(0)=1+\beta_{-}(0)$ are
needed. For the Ohmic bath, and in the limit $\omega_{c}\gg{\rm max}\{ N\gamma,\omega_{00}\}$,
we have found (with $\delta\equiv N\gamma/\omega_{00}$)

\begin{eqnarray}
\alpha(0) & = & -L_{1}+\left(\frac{\omega_{0}}{\omega_{00}}+\frac{\omega_{00}}{\omega_{0}}\left(\frac{\delta^{2}}{2}-1\right)\right)L_{2}\\
\beta_{-}(0) & = & L_{1}+\left(\frac{\omega_{0}}{\omega_{00}}-\frac{\omega_{00}}{\omega_{0}}\left(\frac{\delta^{2}}{2}-1\right)\right)L_{2}-\frac{1}{2}\end{eqnarray}
where $L_{1}=\frac{N\gamma}{2\pi\omega_{0}}\ln\frac{\omega_{c}}{\omega_{0}}$
and

\begin{equation}
L_{2}=\frac{\ln\left[\delta^{2}/2-1+\delta\sqrt{\delta^{2}/4-1}\right]}{2\pi\sqrt{\delta^{2}-4}}\,.\end{equation}
The behaviour of the density matrix is shown in Fig. \ref{cap:Equilibrium-density-matrix:}.
Interestingly, the same form of the equilibrium density matrix has
been obtained in a model of interacting fermions inside a harmonic
oscillator\cite{wonneberger}, with a special interaction, and without
bath. In particular, Eq. (43) of Ref. \onlinecite{wonneberger} may
be compared\cite{footnoteWB} directly to our Eq. (\ref{GreensFunctExt})
for the Green's function, evaluated at $t=0$. Naturally the dynamics
of the two models is different. 

As evident in the plot, there is a 'jump' in the distribution at $n=n_{F}$,
which is given by:

\begin{equation}
\delta\rho=\int_{-\pi}^{\pi}\frac{dx}{2\pi}e^{-2\beta_{-}(0)(1-\cos(x))}I_{0}[2\alpha(0)(\cos(x)-1)]\,.\end{equation}
In the strong-coupling limit, the shape of the distribution is approximated
by a 'discrete error function', 

\begin{equation}
\rho_{nn}\approx\frac{1}{2\sqrt{\pi\beta_{-}(0)}}\sum_{k=n-n_{F}}^{\infty}e^{-k^{2}/(4\beta_{-}(0))}\,.\end{equation}

\begin{figure}
\begin{center}\includegraphics[%
  height=7cm]{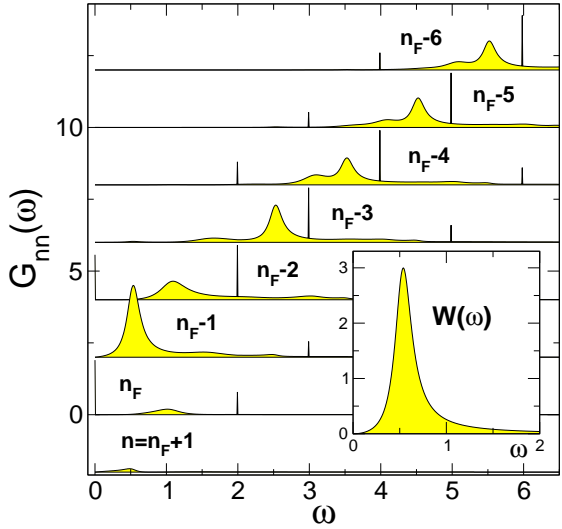}\end{center}

\caption{\label{DFSgreenresult}Green's function of fermions in a damped harmonic
oscillator (with an Ohmic bath): Fourier transform $G_{nn}(\omega)$
of $\left\langle \hat{c}_{n}^{\dagger}(t)\hat{c}_{n}\right\rangle $
plotted vs. $\omega$, for different values of $n$ (curves displaced
vertically for clarity). Note that $\omega$ is measured with respect
to $-n_{F}\omega_{0}$ and we have set $\omega_{0}=1$. The strength
of the coupling is given by $N\gamma/\omega_{00}=0.4$. Inset: Weight
function $W(\omega)$ vs. $\omega$. The height of the smaller $\delta$
peaks is an indication of their weight. }
\end{figure}

\begin{figure}
\begin{center}\includegraphics[%
  height=7cm]{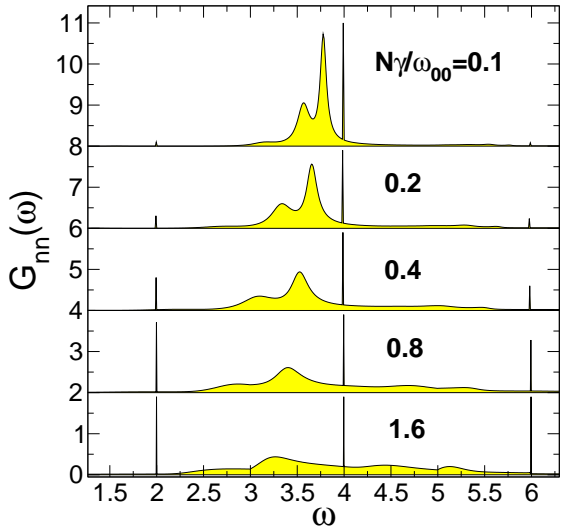}\end{center}

\caption{\label{DFSgreengammas}Dependence of Green's function on coupling
strength: Fourier transform $G_{nn}(\omega)$ of $\left\langle \hat{c}_{n}^{\dagger}(t)\hat{c}_{n}\right\rangle $
vs. $\omega$, for different values of the coupling strength, indicated
in the plot (curves displaced vertically for clarity). Note that $\omega$
is measured with respect to $-n_{F}\omega_{0}$ and plotted in units
of $\omega_{0}$, and that $\omega_{0}$ changes with $\gamma$, for
fixed $\omega_{00}$, see Eq. (\ref{DFSohmbathomega}). This plot
is produced for a fixed excitation number, $n=n_{F}-4$. }
\end{figure}

Regarding the Green's function, the series expansion of Eq. (\ref{DFSg0E})
has been carried out using a symbolic computer algebra system. In
order to obtain the Green's function in frequency space, the resulting
terms of the form $\alpha(t)^{n_{\alpha}}\beta_{+}(t)^{n_{+}}\beta_{-}(t)^{n_{-}}$
have then been Fourier-transformed (which leads to repeated convolutions
of their Fourier spectra, i.e. essentially $W(\omega)$, see Eqs.
(\ref{DFSalpha}),(\ref{DFSbetaplus}),(\ref{DFSbetaminus})). This
is done numerically (using a fast Fourier transform). The results
obtained for the Green's function in the case of coupling to an Ohmic
bath are depicted in Figs. \ref{DFSgreenresult} and \ref{DFSgreengammas}.
(In the expansion of the exponential, the maximum combined power of
$\alpha,\beta_{\pm}$ that was kept has been $6$ for Fig. \ref{DFSgreenresult}
and 10 for Fig. \ref{DFSgreengammas})

The shape of the broadened peaks in the incoherent background reflects
the Fourier-transforms of $\alpha(t),\beta_{\pm}(t)$. All of them
are related to the spectral function $W(\omega$) describing the damped
c.m. motion (see Eqs. \ref{DFSalpha}, \ref{DFSbetaplus} and \ref{DFSbetaminus}).
In particular, at weak coupling, the Fourier transform of the dominant
contribution $\beta_{+}(t)$ is essentially given by $W(\omega)$,
see Eq. (\ref{DFSbetaplus}). 

For the Ohmic bath, whose power spectrum is given in Eq. (\ref{DFSohmicbath}),
we have (see (\ref{DFSfreqbare})):

\begin{equation}
\omega_{0}^{2}=\omega_{00}^{2}+\frac{2}{\pi}N\gamma\omega_{c}\,.\label{DFSohmbathomega}\end{equation}
The resulting spectral function $W(\omega)$ associated to the damped
c.m. motion, see Eqs. (\ref{DFSWdef}) and (\ref{DFSWresult}), is
found to be 

\begin{eqnarray}
 &  & W(\omega)=\frac{2\gamma N}{\pi}\omega^{2}\times\label{DFSohmW}\\
 &  & \left[\left(\omega^{2}-\omega_{00}^{2}-\frac{\gamma N}{\pi}\omega\ln\frac{\omega_{c}+\omega}{|\omega_{c}-\omega|}\right)^{2}+(\gamma N\omega)^{2}\right]^{-1}\,.\nonumber \end{eqnarray}
This expression holds for $0\leq\omega<\omega_{c}$. (The denominator
becomes $0$ at some $\omega>\omega_{c}$, which yields a $\delta$
peak whose weight vanishes for $\gamma N\rightarrow0$. In order to
obtain this peak, one has to replace $2\gamma N\omega$ by an infinitesimal
$+0$ for $\omega>\omega_{c}$. This would be different for smooth
cutoffs, and it does not affect any of our results). The function
$W(\omega)$ is displayed in the inset of Fig. \ref{DFSgreenresult}
(where $\omega_{c}/\omega_{00}=10$). At low frequencies $\omega\ll\omega_{c}$
and damping rate $\gamma N\ll\omega_{c}$, $W(\omega)$ is proportional
to the magnitude squared of the susceptibility derived for the linear
response of the velocity of the damped c.m. oscillator, with renormalized
frequency $\omega_{00}$ and damping rate $\gamma N$:

\begin{equation}
\left|\chi_{vx}(\omega)\right|^{2}\propto\omega^{2}\left[(\omega^{2}-\omega_{00}^{2})^{2}+(\gamma N\omega)^{2}\right]^{-1}\,.\label{DFSchisq}\end{equation}
In the overdamped case, $\gamma N>2\omega_{00}$, the zeroes of the
denominator occur at purely imaginary frequencies. However, the shape
of the function (\ref{DFSchisq}) shows a qualitative change already
at $\gamma N=\sqrt{2}\omega_{00}$, when the curvature at $\omega=0$
changes sign (in the susceptibility the maximum vanishes). This may
be observed by comparing the lowest curve in Fig. \ref{DFSgreengammas},
corresponding to $\gamma N/\omega_{00}=1.6$, to the other curves.

Apart from the generic exponential-type decay at intermediate times,
we can also evaluate the behaviour of $\alpha(t),\,\beta_{+}(t)$
and $\beta_{-}(t)$ at long times $t\gg1/{\rm min}(\omega_{00},N\gamma)$,
for the special case of the Ohmic bath. All three functions then decay
like $t^{-2}$, namely as $\alpha(t)\approx\beta_{\pm}(t)\approx-[N\gamma\omega_{0}/(2\pi\omega_{00}^{4})]/t^{2}$,
which corresponds to the decay of the incoherent part of the Green's
function.

The \emph{general} long-time behaviour of the Green's function (independently
of the bath spectrum) can be read off directly from the exact expression
given above in Eq. (\ref{DFSg0E}): Since the coupling to the bath
damps away the c.m. motion, the functions $\alpha(t)$, $\beta_{+}(t)$
and $\beta_{-}(t)$ all decay to zero at $t\rightarrow\infty$ (excluding
cases such as a gapped bath). The exponent $\delta E$ in Eq. (\ref{DFSexponent})
thus approaches a finite value. This leads to the conclusion that
the Green's function $\left\langle \hat{c}_{n_{F}-\delta n}^{\dagger}(t)\hat{c}_{n_{F}-\delta n}\right\rangle $
does not decay to zero in the limit $t\rightarrow\infty$, for any
$\delta n>1$, since contributions from $(X'/X)^{\delta n}$ with
$\delta n>1$ remain. In frequency space, it consists of a number
of delta peaks superimposed onto an incoherent background related
to the damped c.m. mode (see plots). These facts can also be seen
from the simple weak-coupling expression, Eq. (\ref{weakcouplGF}).

This result may be unexpected, since according to a \emph{naive} application
of the Golden Rule to the Hamiltonian (\ref{DFSham}), any electron
inserted into the system at $n>n_{F}+1$ (and any hole inserted at
$n<n_{F}$) should decay towards lower (higher) single-particle energy
levels by spontaneous emission of energy into the bath (for $T=0$).
The rate is expected to be:

\begin{equation}
\Gamma_{GR}=2\pi\left\langle \hat{F}\hat{F}\right\rangle _{\omega_{0}}\frac{n_{F}}{2m\omega_{0}}\,.\end{equation}
(with $n\approx n_{F}\approx N$). For an electron inserted just above
the Fermi surface ($n=n_{F}+1$), or a hole inserted at $n=n_{F}$,
the Pauli principle blocks relaxation and the corresponding Golden
Rule rate vanishes. The decay rate of Green's functions is expected
to be $\Gamma_{GR}/2$. In the following, we discuss the deviations
from this naive expectation.

For weak coupling, the decay of the Green's function $\left\langle \hat{c}_{n}^{\dagger}(t)\hat{c}_{n}\right\rangle $
calculated above is determined by the decay of $\beta_{+}(t)$, while
$\alpha$ and $\beta_{-}$ are approximately zero. For example, at
$n=n_{F}-1$ the hole-propagator is directly proportional to $\beta_{+}(t)$:

\begin{equation}
\left\langle \hat{c}_{n_{F}-1}^{\dagger}(t)\hat{c}_{n_{F}-1}\right\rangle \approx\beta_{+}(t)e^{in_{F}\omega_{0}t}\,.\end{equation}
 The decay of this function is determined by the width of the Lorentz
peak appearing at $\omega\approx\omega_{0}$ in the {}``spectral
weight'' $W(\omega)$ defined above (see Eqs. (\ref{DFSWdef}) and
(\ref{DFSbetaplus})) and calculated in Appendix \ref{DFSappendixDiagon}.
According to Eqs. (\ref{DFSWresult}) and (\ref{DFSappGammaDef}),
at weak coupling this width is given by $\Delta\omega=\Gamma_{GR}/2$,
as expected. Therefore, there is quantitative agreement between the
exact calculation and the Golden Rule result at $n=n_{F}-1$. This
is true also at $n=n_{F}$, where the Green's function evaluated in
the preceding section only has a delta peak, related to Pauli blocking
(apart from corrections that vanish in the limit of weak coupling),
see Fig. \ref{DFSgreenresult}. However, at $n<n_{F}-1$, there is
always an additional $\delta$ peak that is incompatible with the
Golden Rule expectation, as explained above. Essentially, this discrepancy
stems from the fact that the many-particle states in the harmonic
oscillator are in general strongly degenerate. Therefore, the preconditions
for the usual derivation of a master equation \cite{Blum} are violated.
We will trace back this behaviour to the fact that the exact wave
function splits into c.m. and relative motion (see the remainder of
this section and the following section). Readers not interested in
this detailed analysis may wish to skip directly to Section \ref{DFStwoparticlemaintext},
for the time-evolution of the density matrix.

The excited many-particle state $\kt{\Psi}$, which is created by
adding an extra hole (or extra particle) and whose time-evolution
determines the Green's function, can be written as a superposition
of states, each of which distributes the given number of excitation
quanta $\delta n$ differently onto center-of-mass ({}``cm'') and
relative ({}``r'') motion. Therefore, in the weak-coupling limit,
where one starts out of the factorized ground state $\kt{0}_{cm}\kt{0}_{r}\kt{0}_{B}$,
it is of the form:

\begin{equation}
\kt{\Psi}=\kt{0}_{B}(a_{0}\kt{0}_{cm}\kt{\delta n}_{r}+a_{1}\kt{1}_{cm}\kt{\delta n-1}_{r}+\ldots)\,.\label{DFSpsicmr}\end{equation}
The first state does not contain any extra excitation of the c.m.
mode. Therefore, it is unaffected by the c.m.-bath coupling and does
not decay. This leads to the main $\delta$-peak contribution at a
frequency $\delta n\omega_{0}$, which is determined by the number
of excitation quanta $\delta n$. 

The other states will decay. The resulting broadened peaks are shifted
towards lower frequencies, due to the renormalization of the c.m.
mode frequency to the value $\omega_{00}$. Thus these peaks occur
at frequencies of the form $(\delta n-n_{cm})\omega_{0}+n_{cm}\omega_{00}$,
where $n_{cm}$ counts the number of quanta in the c.m. mode. This
is visible in Figs. \ref{DFSgreenresult} and \ref{DFSgreengammas},
where the Fourier transform of the Green's function is displayed.
In the limit of weak coupling ($\alpha=\beta_{-}=0$ and $\beta_{+}(0)=1$
), the expansion yields the following weights of the different peaks:

\begin{equation}
\left|a_{n_{cm}}\right|^{2}=\frac{1}{n_{cm}!}\sum_{j=0}^{\delta n-n_{cm}}\frac{(-1)^{j}}{j!}\,,\label{DFSweightcm}\end{equation}
for $0\leq n_{cm}\leq\delta n$. These weights are normalized ($\sum_{n_{cm}=0}^{\delta n}\left|a_{n_{cm}}\right|^{2}=1$),
since the application of a particle annihilation operator onto the
non-interacting ground state creates a normalized state $\kt{\Psi}$
(for $\delta n\geq0$). In particular, the weight of the $n_{cm}=0$-component,
which leads to the $\delta$ peak, goes to the constant value of $1/e$
in the limit $\delta n\rightarrow\infty$. Thus, there is a non-decaying
contribution even at arbitrarily high excitation energies. This fact
is connected to the assumption of large $N$ under which the present
results have been obtained (see the following section). The weight
of the contribution from $n_{cm}=\delta n-1$ always vanishes, as
may be observed in the plots as well. 

For stronger coupling (Fig. \ref{DFSgreengammas}), other $\delta$
peaks start to grow both above and below the frequency of the main
peak. These arise because the interacting ground state contains contributions
from excited c.m. states and bath states. We can write it down in
the following schematic form, where $\kt{j}_{B}$ denotes a bath state
for which the sum of all harmonic oscillator excitations is equal
to $j$:

\begin{eqnarray}
 &  & \kt{GS}_{cm+B}=\label{DFSGScmB}\\
 &  & c_{00}\kt{0}_{cm}\kt{0}_{B}+c_{11}\kt{1}_{cm}\kt{1}_{B}+c_{20}\kt{2}_{cm}\kt{0}_{B}+\ldots\,.\nonumber \end{eqnarray}
Any of the excited states discussed above, containing $n_{cm}$ c.m.
excitations (see Eq. (\ref{DFSpsicmr})), may therefore have a finite
overlap with this ground state of the coupled system c.m.-bath. Thus,
a non-decaying contribution at the frequency $(\delta n-n_{cm})\omega_{0}$
of the remaining excitation in the \emph{relative} motion survives.
This explains the $\delta$ peaks at frequencies equal to integer
multiples of $\omega_{0}$, situated \emph{below} the unperturbed
excitation energy of $\delta n\omega_{0}$. There are also peaks at
still \emph{larger} frequencies, because the additional particle (or
hole) is introduced into the many-particle ground state that already
contains excitations of the c.m. mode, see Eq. (\ref{DFSGScmB}).
Therefore, the resulting excited many-particle state $\kt{\Psi}$
will also contain contributions with more than $\delta n$ excitations
of c.m. and relative motion combined, in contrast to the weak-coupling
form, Eq. (\ref{DFSpsicmr}). Since adding $\delta n$ excitations
to one of the states $\kt{j}_{cm}\kt{0}_{r}$ that appear in the many-particle
ground state may lead to states containing up to $\delta n+j$ excitations
in the \emph{relative} motion alone, this explains the appearance
of $\delta$ peaks at $(\delta n+j)\omega_{0}$ in the Green's function.
For the same reason, the {}``incoherent background'' due to the
c.m. mode extends beyond the main frequency of $\delta n\omega_{0}$.

Finally, it may be noted from the figures that the additional $\delta$
peaks appear only at frequencies of the form $(\delta n+2j)\omega_{0}$,
i.e. they are removed by an \emph{even} multiple of $\omega_{0}$
from the main peak. Physically, the reason is the following: The coupled
ground state of c.m. and bath remains symmetric under inversion of
the c.m. and oscillator coordinates. Therefore, the sum of c.m. and
bath excitations is always even, as indicated in (\ref{DFSGScmB}).
When adding $\delta n$ extra excitations to the many-particle ground
state (which is (\ref{DFSGScmB}), multiplied by the ground state
of the relative motion), a part of those excitations may go into the
c.m. motion, leading to peaks below and above the main frequency.
Only if the number of added or subtracted c.m. excitations is even,
a nonvanishing overlap with the many-particle ground state may develop,
leading to a non-decaying component in the Green's function.

The considerations of the current section apply to the approximate
Hamiltonian (\ref{DFShapprox}) that is good in the limit $N\rightarrow\infty$
of large particle number, and which has been amenable to bosonization.
However, the qualitative arguments concerning the contributions of
c.m. and relative motion remain valid for finite $N$ as well. This
is the topic of section \ref{sec:Finite-particle-number:}.

\section{Finite particle number: Center-of-mass motion in excited Fock states}

\label{sec:Finite-particle-number:}In this section we analyze in
more detail the splitting into c.m. and relative motion for arbitrary
\emph{finite} $N$. We will explain that the weight of the coherent
$\delta$ peak in the Green's function decays when moving towards
higher excitation energies, on a scale set by the number $N$ of particles.
This is why, in the limit $N\rightarrow\infty$ considered for the
bosonized model, this weight even saturates at a constant value for
excited states arbitrarily far above the Fermi surface. We will not
actually evaluate Green's functions for the finite $N$ case, which
is still much more difficult.

For brevity, we set $m=\omega_{0}=1$ throughout this section. The
coordinates and momenta of the individual particles are related to
the harmonic oscillator raising and lowering operators by

\begin{equation}
\hat{x}_{j}=\frac{1}{\sqrt{2}}(\hat{a}_{j}+\hat{a}_{j}^{\dagger}),\,\hat{p}_{j}=\frac{-i}{\sqrt{2}}(\hat{a}_{j}-\hat{a}_{j}^{\dagger})\,,\end{equation}
where it is understood that $\hat{a}_{j}$ acts only on the coordinate
of particle $j$. The center-of-mass motion is described by

\begin{equation}
\hat{X}=\frac{1}{N}\sum_{j=1}^{N}\hat{x}_{j},\,\hat{P}=\sum_{j=1}^{N}\hat{p}_{j}\,,\end{equation}
such that we obtain the following operator that lowers the excitation
of the c.m. harmonic oscillator by one:

\begin{equation}
\hat{A}=\sqrt{\frac{N}{2}}(\hat{X}+i\frac{\hat{P}}{N})=\frac{1}{\sqrt{N}}\sum_{j=1}^{N}\hat{a}_{j}\,.\end{equation}
Here we have used that the total mass is $M=Nm=N$.

Let us consider an $N$-particle Fock state $\kt{\Psi_{N,\delta n}}$
consisting of the Fermi sea filled up to (and including) level $n_{F}-1=N-2$
and an additional single particle that has been placed into the excited
level at $n_{F}+\delta n$. It is our goal to find the probability
$P_{n}$ of having $n$ excitations in the c.m. mode, given this many-particle
state. In particular, the probability of having $0$ excitations will
be the weight of the {}``coherent component'' that does not decay
in spite of coupling to the bath. This holds quantitatively in the
weak-coupling limit, where we may neglect the change in the eigenstates
of the c.m. mode, see the previous discussion. For arbitrary coupling,
it is still the correct qualitative picture. 

First, we rewrite $\hat{A}$ in second quantization:

\begin{equation}
\hat{A}=\frac{1}{\sqrt{N}}\sum_{k=0}^{\infty}\sqrt{k+1}\hat{c}_{k}^{\dagger}\hat{c}_{k+1}\,.\end{equation}
Using this, it is straightforward to check that 

\begin{equation}
\left\langle \Psi_{N,\delta n}|\hat{A}^{\dagger j}\hat{A}^{j}|\Psi_{N,\delta n}\right\rangle =\frac{\delta n+n_{F}}{N}\cdot\ldots\cdot\frac{\delta n+n_{F}-j+1}{N}\,,\label{DFSajajfock}\end{equation}
for $j\leq\delta n$, otherwise this is $0$. On the other hand, if
we think of $\kt{\Psi_{N,\delta n}}$ as being written in a basis
that splits relative and c.m. motion (compare Eq. (\ref{DFSpsicmr}))
and take into account the usual matrix elements of a harmonic oscillator
lowering operator, we obtain for the same expectation value (with
$P_{n}$ the occupation probabilities for the different c.m. states):

\begin{equation}
\sum_{n=j}^{\infty}P_{n}n(n-1)\cdot\ldots\cdot(n-j+1)\,.\label{DFSajajoscillator}\end{equation}
Now we make use of the fact that the total number of excitations in
the c.m. mode is limited by $\delta n$, i.e. $P_{n}=0$ for $n>\delta n$.
Therefore, it is possible to start at $j=\delta n$, equate (\ref{DFSajajfock})
to (\ref{DFSajajoscillator}), solve for $P_{\delta n}$, and then
proceed iteratively all the way down to $P_{0}$. In each step, only
the probabilities that have been calculated before appear in addition
to the unknown $P_{j}$. We can write down the set of equations very
transparently by introducing the abbreviation

\begin{equation}
p_{l}=(\delta n-l)!\frac{(N-1)!}{(\delta n+N-1)!}N^{\delta n}\, P_{\delta n-l}\,.\label{DFSdjabbrev}\end{equation}
Then we have (from equating (\ref{DFSajajfock}) to (\ref{DFSajajoscillator}))
the recursive relation:

\begin{equation}
p_{l}=\frac{N^{l-1}}{(N+l-1)\cdot\ldots\cdot(N+1)}-\sum_{k=1}^{l}\frac{p_{l-k}}{k!}\,.\label{DFSdlrecursive}\end{equation}
The solution starts with $p_{0}=1$. Note that we always get $p_{1}=0$
(for any $N$), corresponding to $P_{\delta n-1}=0$: This has been
observed already in the context of the bosonization solution (see
previous section). It is also remarkable that the equation for $p_{l}$
does not depend on $\delta n$. Therefore, we can solve the problem
at once for arbitrary $\delta n$ (which then only enters in $P_{n}$
via Eq. (\ref{DFSdjabbrev})). %
\begin{figure}
\begin{center}\includegraphics[%
  height=6cm]{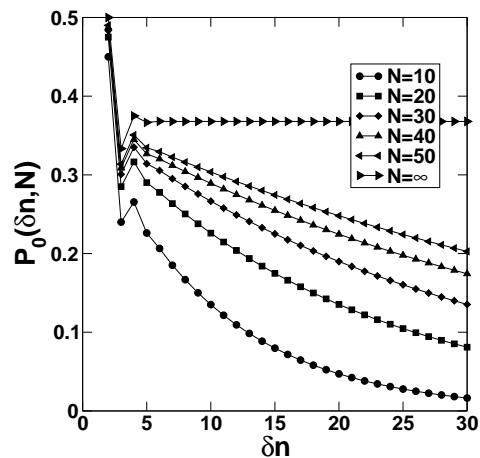}\end{center}

\caption{\label{DFScmweights}Weight $P_{0}$ of the center-of-mass ground
state (Eq. (\ref{P0eq})) as a function of excitation $\delta n$,
for different particle numbers $N$, including the bosonization result
from Eq. (\ref{DFSweightcm}) ({}``$N=\infty$'').}
\end{figure}
After solving these equations, it is found that the weight of the
c.m. ground state, 

\begin{equation}
P_{0}=\frac{(\delta n+N-1)!}{(N-1)!}N^{-\delta n}\, p_{\delta n}\,,\label{P0eq}\end{equation}
decays as a function of excitation energy $\delta n$ (for large $\delta n$),
on a scale set by $N$, see Fig. \ref{DFScmweights}.

Finally, we remark that the general structure of the exact eigenfunctions
of our problem (defined by the original Hamiltonian, Eq. (\ref{DFSham}))
remains identical to that found in the limit $N\rightarrow\infty$,
using bosonization. For arbitrary system-bath coupling, the eigenstates
of the full system will be of the form

\begin{equation}
\kt{\Psi^{SB}}=\kt{\Phi^{CM-B}}\otimes\kt{\psi^{rel}}\,,\end{equation}
where {}``SB'' refers to the full Hilbert space of fermions coupled
to a bath, while $\kt{\Phi^{CM-B}}$ is a new eigenstate of the coupled
system c.m./bath and $\kt{\psi^{rel}}$ is an \emph{unaltered}, antisymmetric
eigenstate of the relative motion. However, the evaluation of quantities
like Green's functions or the reduced density matrix, starting from
this expression, becomes very cumbersome, which is the reason why
we have restricted most of the discussion to the bosonization solution.

\section{Two-particle Green's function: decay of populations and dephasing}

\label{DFStwoparticlemaintext}While the single-particle Green's function
reveals how fast an electron is scattered out of an initial state,
we have to turn to the two-particle Green's function in order to learn
about the time-evolution of the reduced single particle density matrix.
This describes relaxation and dephasing. For our purposes, we will
be particularly interested in the following two-particle Green's function
of fermions in the damped oscillator (the evaluation of other four-point
correlators proceeds analogously):

\begin{equation}
\left\langle \hat{c}_{n}(0)\hat{c}_{l'}^{\dagger}(t)\hat{c}_{l}(t)\hat{c}_{n'}^{\dagger}(0)\right\rangle \,.\label{DFS2pgfwewant}\end{equation}
Without coupling to the bath, we could apply Wick's theorem to obtain:

\begin{eqnarray}
\left\langle \hat{c}_{n}(0)\hat{c}_{l'}^{\dagger}(t)\hat{c}_{l}(t)\hat{c}_{n'}^{\dagger}(0)\right\rangle =\nonumber \\
\delta_{nl'}\delta_{ln'}(1-\left\langle \hat{n}_{n}\right\rangle )(1-\left\langle \hat{n}_{n'}\right\rangle )e^{i\omega_{0}(n-n')t}+\nonumber \\
\delta_{nn'}\delta_{ll'}(1-\left\langle \hat{n}_{n}\right\rangle )\left\langle \hat{n}_{l}\right\rangle \,,\label{DFSwick}\end{eqnarray}
If the bath damps the motion of the fermions, we may rewrite (\ref{DFS2pgfwewant})
in terms of the auxiliary fermion operators $\hat{\psi}(x)$ defined
in (\ref{DFScpsi}):

\begin{eqnarray}
\left\langle \hat{c}_{n}(0)\hat{c}_{l'}^{\dagger}(t)\hat{c}_{l}(t)\hat{c}_{n'}^{\dagger}(0)\right\rangle =\nonumber \\
\frac{1}{(2\pi)^{2}}\int_{0}^{2\pi}dxdx'dydy'\, e^{i(l'x'+n'y'-lx-ny)}\times\nonumber \\
\left\langle \hat{\psi}(y,0)\hat{\psi}^{\dagger}(x',t)\hat{\psi}(x,t)\hat{\psi}^{\dagger}(y',0)\right\rangle \label{DFS2pgfviapsi}\end{eqnarray}
Starting from the representation of $\hat{\psi}$ in terms of boson
operators $\ph$, Eq. (\ref{DFSpsibos}), this may be evaluated using
the same methods as in Sections \ref{bosonAndDiagon} and \ref{DFSgreensfctT0}.
In Appendix \ref{DFStwoparticleappendix}, we show how the two-particle
Green's function is obtained via a series expansion, similar to the
single-particle Green's function. The general exact result is given
in Eqs. (\ref{DFSa0etilde})-(\ref{DFSetilde}). Although in principle
this expression permits evaluation of the two-particle Green's function
at arbitrary coupling strength, this task has proven to be computationally
much more difficult than the analogous calculation for the single-particle
Green's function, due to the larger number of terms generated in the
power series. Nevertheless, the weak-coupling results already contain
many nontrivial features. 

We may now answer questions like the following: If we introduce an
extra single particle in some level $\tilde{n}$ above the Fermi sea,
how does it decay towards lower-lying levels, by emitting some of
its energy into the bath? We first look at the time-evolution of the
populations

\begin{equation}
\rho_{nn}(t)=\left\langle \hat{c}_{\tilde{n}}\hat{c}_{n}^{\dagger}(t)\hat{c}_{n}(t)\hat{c}_{\tilde{n}}^{\dagger}\right\rangle \,,\label{DFSpop}\end{equation}
where the two outermost operators create the desired state at time
$0$, while the two operators in the middle test for the population.
We note that, in principle, the newly created state may have norm
less than $1$, if the level $\tilde{n}$ happens to be partially
occupied already in the ground state on which the creation operator
$\hat{c}_{\tilde{n}}^{\dagger}$ acts. In that case, it would be necessary
to divide (\ref{DFSpop}) by the norm, $\left\langle \hat{c}_{\tilde{n}}\hat{c}_{\tilde{n}}^{\dagger}\right\rangle $.
However, since we will restrict the evaluation to the weak-coupling
case, the states above $n_{F}$ are initially empty (at $T=0$). 

\begin{figure}
\begin{center}\includegraphics[%
  width=0.95\columnwidth]{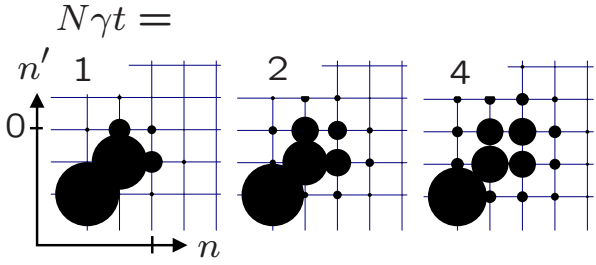}\end{center}

\caption{\label{FIGrhoheat}Time-evolution of $\rho_{{\rm heat}}(n,n',t)$,
describing heating around the Fermi surface due to an extra particle
placed in a highly excited state above the Fermi sea (Eq. (\ref{rhoheat})).
See text for connection to two-particle Green's function, Eq. (\ref{rhoSimple}).
In this plot $\omega_{0}=\omega_{0}'$. The radius of each dot gives
$\left|\rho_{nn'}(t)\right|$ (being $1$ for the two occupied states
in the lower left corner). }
\end{figure}

\begin{figure}
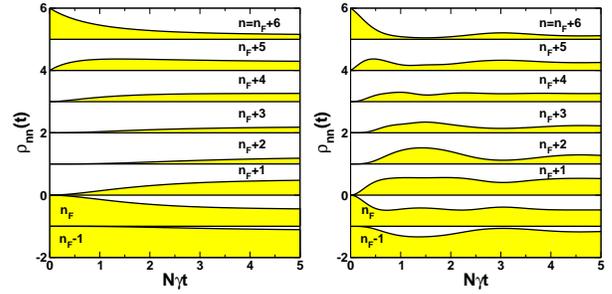

\begin{center}\includegraphics[%
  width=0.45\columnwidth,
  height=7cm,
  keepaspectratio]{FINALpopDecay.eps}~\includegraphics[%
  width=0.45\columnwidth,
  height=7cm,
  keepaspectratio]{FINALpopDecayWiggly.eps}\end{center}

\caption{\label{DFSpopulationsdecay}Time-evolution of the populations $\rho_{nn}(t)$
(Eq. (\ref{DFSpop})) for an electron placed in an initial state $\tilde{n}=n_{F}+6$
above the Fermi sea (weak-coupling result) with $\omega_{0}=\omega_{0}'$
(left) or $\omega_{0}=\omega_{0}'+2N\gamma$ (right), see text. The
initial and final states are the same in both cases. Different curves
have been displaced vertically for clarity. Note the incomplete decay
and heating around the Fermi surface.}
\end{figure}

\begin{figure}
\begin{center}\includegraphics[%
  width=0.90\columnwidth,
  height=7cm,
  keepaspectratio]{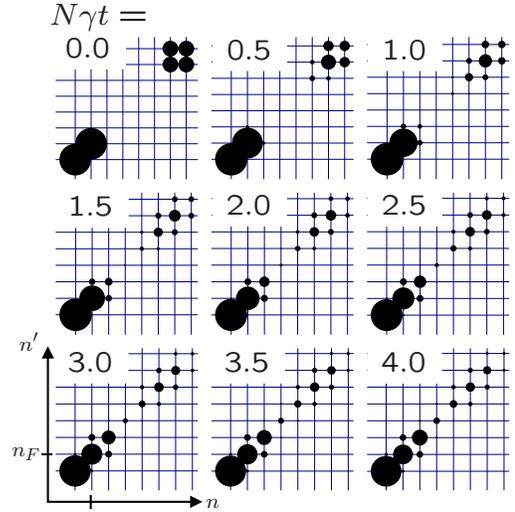}\end{center}

\caption{\label{DFSsuperpositiondecay}Time-evolution of the single-particle
density matrix $\rho_{nn'}(t)$ (Eq. (\ref{DFSrhosup})) starting
from an extra particle in a superposition of states $n_{1}=5,\, n_{2}=6$
above the Fermi sea ($\omega_{0}=\omega_{0}'$). Levels $n,n'=n_{F}-1,n_{F},\ldots,n_{F}+6$
are indicated by grid lines. }
\end{figure}

In particular, we have been able to derive simplified expressions
in the limit of weak-coupling and high initial excitation energy (see
Appendix \ref{sec:Limiting-expressions-for}). We obtain:

\begin{eqnarray}
 &  & \left\langle \hat{c}_{n}(0)\hat{c}_{l'}^{\dagger}(t)\hat{c}_{l}(t)\hat{c}_{n'}^{\dagger}(0)\right\rangle \approx e^{i\omega_{0}(n-n')t}\delta_{n'-l,n-l'}\times\label{rhoSimple}\\
 &  & \left\{ \rho_{{\rm decay}}(n'-l,t)+\rho_{{\rm heat}}(l-n_{F}-1,l'-n_{F}-1,t)\right\} \nonumber \end{eqnarray}
with the first part describing the decay of the excitation,

\begin{equation}
\rho_{{\rm decay}}(m,t)=\frac{(-1)^{m}}{m!}\left(\nu(t)+\nu^{*}(t)\right)^{m}e^{\nu(t)+\nu^{*}(t)}\,,\label{rhodecay}\end{equation}
where $m=n_{0}-n=n_{0}'-n'$ may be interpreted as the net number
of quanta transferred to the bath ($\nu(t)$ is defined in Eq. (\ref{nu})).
The other part describes {}``heating'' around the Fermi surface
(see below):

\begin{equation}
\rho_{{\rm heat}}(n,n',t)=\nu(t)^{n-n'}\sum\frac{\left|\nu(t)\right|^{2(\tilde{m}_{1}+\tilde{m}_{2})}}{m_{1}!m_{2}!\tilde{m}_{1}!\tilde{m}_{2}!}\,(-1)^{m_{2}+\tilde{m}_{2}},\label{rhoheat}\end{equation}
where the triple sum runs over $\tilde{m}_{1}={\rm max}(0,n'+1)\ldots\infty,$
$\tilde{m}_{2}=0\ldots\infty$, $m_{1}={\rm max}(0,\tilde{m}_{2}+n+1)\ldots n-n'+\tilde{m}_{1}+\tilde{m}_{2}$,
and we have $m_{2}=\tilde{m}_{1}+\tilde{m}_{2}-m_{1}+n-n'$. We note
that the limiting case (\ref{rhoSimple}) appears to be a very good
approximation to the full result, even for small excitation energies.
The behaviour of $\rho_{nn}(t)$ shown in Fig. \ref{DFSpopulationsdecay}
gives an impression of $\rho_{{\rm decay}}(m,t)$ for the levels not
directly near the Fermi surface (with $m$ being the distance from
the initial excited state). A plot of $\rho_{{\rm decay}}$ can be
found in Ref. \onlinecite{FlorianDima}. In Fig. \ref{FIGrhoheat},
we have shown the time-evolution of $\rho_{{\rm heat}}$, which results
in a good approximation to the full evolution displayed in Fig. \ref{DFSsuperpositiondecay},
taking into account Eq. (\ref{rhoSimple}). 

The time-evolution of the populations, $\rho_{nn}(t)$, is shown in
Fig. \ref{DFSpopulationsdecay}, which has been calculated in the
weak-coupling limit, but without the approximation of a high initial
excitation. Note that, in general, the difference between $\omega_{0}$
and $\omega'_{0}$ can be of any sign, depending on the details of
the bath spectrum. This frequency shift between the c.m. mode and
all the other boson modes may lead to beating oscillations that may
be visible in the time-evolution of the populations or the density
matrix, which is shown in Fig. \ref{DFSpopulationsdecay}, right.
In Figs. \ref{DFSpopulationsdecay} (left) and \ref{DFSsuperpositiondecay},
however, we have assumed $\omega'_{0}=\omega_{0}$ (e.g. the bath
spectrum is symmetric around the transition frequency, for a fixed
decay rate of $N\gamma$). We note that, contrary to naive expectation,
the particle does not decay all the way down to the lowest unoccupied
state $n_{F}+1$. Rather, in the long-time limit (which is already
reached at $N\gamma t\approx5$ to a good approximation), the extra
particle is distributed over the range of excited levels above the
Fermi surface, up to the initial level $\tilde{n}$. Again, this is
because only the c.m. mode couples to the bath, such that, even at
$T=0$, a fraction of the initial excitation energy remains in the
system. For the limiting case of high excitation energy, discussed
above, this may be seen from $\rho_{{\rm decay}}(m,t)\rightarrow(2^{m}/m!)e^{-2}$.
Moreover, we see that the population of the highest occupied states
in the Fermi sea decreases. The fermions in these states become partly
excited by absorbing energy from the extra particle, due to the effective
interaction mediated by the bath. This is described by $\rho_{{\rm heat}}$
of Eq. (\ref{rhoheat}). 

Another application of the two-particle Green's function (\ref{DFS2pgfviapsi})
consists in looking at the decay of a superposition of states. We
imagine that, at time $0$, a particle is placed in an equally-weighted
superposition of two levels $n_{1}$ and $n_{2}$ above the Fermi
surface. This will introduce off-diagonal contributions in the single-particle
density matrix of the fermions. The subsequent relaxation will suppress
the coherence (i.e. the off-diagonal elements) and transfer population
to the lower levels. In Fig. \ref{DFSsuperpositiondecay}, we have
plotted the time-evolution of the density matrix defined by

\begin{equation}
\rho_{nn'}(t)=\frac{1}{2}\left\langle (\hat{c}_{n_{1}}+\hat{c}_{n_{2}})\hat{c}_{n'}^{\dagger}(t)\hat{c}_{n}(t)(\hat{c}_{n_{1}}^{\dagger}+\hat{c}_{n_{2}}^{\dagger})\right\rangle \,.\label{DFSrhosup}\end{equation}
Again, we have evaluated only the weak-coupling case, such that no
additional normalization factor is necessary (for $n_{1},n_{2}>n_{F}$).

Apart from the features already mentioned above, we observe the decay
of the off-diagonal elements (i.e. dephasing) to be incomplete. A
part of the coherence survives in the relative motion which is unaffected
by the bath. This is also evident from the limit $\rho_{{\rm decay}}(m,t\rightarrow\infty)$
discussed above. If a slight anharmonicity were introduced in the
potential, such that c.m. and relative motion are no longer independent,
we would expect to see the same behaviour at short to intermediate
times. Only in the long-time limit, the fermion system would relax
fully to the $N+1$-particle Fermi sea ground state.

\section{Relaxation with arbitrary energy transfer}

\label{DFSrelaxarbitraryenergytransfer}Up to now, we have considered
a coupling between system and bath where the particle coordinates
enter linearly (since this corresponds to a direct many-particle generalization
of the well-known Caldeira-Leggett model). The coordinate operator
only couples adjacent oscillator levels. Therefore, a particle introduced
somewhere above the Fermi sea will have to relax in a series of spontaneous
emissions, each of them leading down exactly one level. For this reason,
the influence of Pauli blocking becomes apparent only immediately
above the Fermi level. In addition, this type of coupling also leads
to damping of the center-of-mass motion only, such that the system
is non-ergodic.

In this section, we are going to discuss a more general model, where
the bath may induce transitions between oscillator levels that are
farther apart. The most important consequence will be that the presence
of the Fermi sea leads to a strong dependence of decay rates on the
distance to the Fermi level. As one moves towards the Fermi level,
more and more of the transitions will be blocked due to the Pauli
principle. It is interesting to observe the consequences of this in
an exactly solvable model, where we are able to go beyond a Golden
Rule description. Furthermore, this coupling will (in general) lead
to ergodic behaviour, i.e. full relaxation of the excitation that
has been added to the system.

We start from the bosonized Hamiltonian, Eq. (\ref{DFShapprox}),
and replace the coupling term by a more general form: 

\begin{eqnarray}
\sqrt{\frac{N}{2m\omega_{0}}}\hat{F}\sum_{q=1}^{\infty}f_{q}\,\sum_{n}\left(\hat{c}_{n}^{\dagger}\hat{c}_{n+q}+h.c.\right)=\nonumber \\
\sqrt{\frac{N}{2m\omega_{0}}}\hat{F}\sum_{q=1}^{\infty}f_{q}\sqrt{q}(\hat{b}_{q}+\hat{b}_{q}^{\dagger})\,.\label{DFSnewcoupling}\end{eqnarray}

Now the bath may induce transitions between levels $n+q$ and $n$,
with an arbitrary amplitude $\propto f_{q}$ (which, however, must
not depend on $n$). For $f_{1}=1,\, f_{q}=0\,(q>1)$ we would recover
the original model. In order to simplify the following expressions
somewhat, we have assumed $f_{q}$ to be real-valued (nothing essential
would change for complex amplitudes). For $f_{q}=const$, the amplitude
of transitions between two levels does not depend on their distance.
The overall prefactor has been kept the same only in order to retain
similarity with the previous results (it may always be compensated
by appropriate choice of the amplitudes $f_{q}$). Note that, in terms
of the original model (before bosonization), the interaction given
here only approximately corresponds to a nonlinear coupling that adds
higher powers of the particle coordinates, since such a coupling would
lead to additional terms in the bosonized version. 

We will take the interaction term Eq. (\ref{DFSnewcoupling}) as the
starting point of our analysis, which employs the methods introduced
before. In the weak-coupling limit, we expect each of the boson modes
to be damped with a Golden Rule decay rate that follows directly from
Eq. (\ref{DFSnewcoupling}):

\begin{equation}
\Gamma_{q}=2\pi\,\frac{N}{2m\omega_{0}}f_{q}^{2}q\,\left\langle \hat{F}\hat{F}\right\rangle _{q\omega_{0}}^{T=0}\,.\label{DFSdecayrateq}\end{equation}
Adding a particle above the Fermi sea, in state $n_{F}+\delta n+1$,
will introduce bosonic excitations of up to $q=\delta n$ (in the
weak-coupling limit, starting from the unperturbed system), and its
decay will be governed by corresponding rates $\Gamma_{q}$. As $\delta n$
grows, the typical decay rate will grow as well, as expected. This
will be analyzed in more detail further below.

Now all the boson modes couple to all the bath oscillators. We retain
the definition of oscillator annihilation operators $\hat{d}_{j}$
given in Eq. (\ref{DFSdop}) for the bath modes $j\geq1$, while the
boson modes are now defined to have negative indices, $j=-q\leq-1$,
with $\hat{d}_{-q}\equiv\hat{b}_{q}$ and $\Omega_{-q}\equiv q\omega_{0}\,$
for $q\geq1$ (there is no $j=0$ mode in this notation). Using Eq.
(\ref{DFSdop}), this definition leads to $\sqrt{q}(\hat{b}_{q}+\hat{b}_{q}^{\dagger})=q\sqrt{2m\omega_{0}}\,\hat{Q}_{-q}$.
The resulting problem of coupled oscillators may again be written
in the form of Eq. (\ref{DFShprime}). However, now the {}``perturbation''
matrix $V$ that couples boson modes and bath oscillators has the
following nonvanishing entries ($j,q\geq1$):

\begin{equation}
V_{-q,j}=V_{j,-q}=f_{q}q\cdot\frac{g}{m}\sqrt{\frac{N}{N_{B}}}\,.\label{DFSnewV}\end{equation}
After diagonalizing the problem with an orthogonal matrix $C$ (compare
Appendix \ref{DFStwoparticleappendix}, Eqs. (\ref{DFSdiagonC})-(\ref{DFSdtildedef})),
the required correlators of boson modes, of the type $\left\langle \hat{b}_{q'}(t)\hat{b}_{q}^{\dagger}\right\rangle $,
can be calculated exactly as before. Now the interaction with the
bath also mixes different boson modes. Therefore, we may have $q'\neq q$
in general. The appropriate generalization of the definition for the
spectral weight $W$ (Eq. (\ref{DFSWdef})) is

\begin{equation}
W_{q'q}(\omega)\equiv\sum_{j\neq0}C_{-q',,j}C_{-q,j}\delta(\omega-\tilde{\Omega}_{j})\,.\end{equation}
At $T=0$, we obtain, using Eq. (\ref{DFSdnew}) (in analogy to (\ref{DFSgrfct})
or (\ref{DFSbetaplus})):

\begin{eqnarray}
\left\langle \hat{b}_{q'}(t)\hat{b}_{q}^{\dagger}\right\rangle =\frac{1}{4}\int_{0}^{\infty}\, W_{q'q}(\omega)\, e^{-i\omega t}\,\times\nonumber \\
\left(\sqrt{\frac{q'\omega_{0}}{\omega}}+\sqrt{\frac{\omega}{q'\omega_{0}}}\right)\left(\sqrt{\frac{q\omega_{0}}{\omega}}+\sqrt{\frac{\omega}{q\omega_{0}}}\right)\, d\omega.\label{DFSnewbb}\end{eqnarray}
Analogous expressions hold for the other correlators: The sign in
the first bracket on the r.h.s., involving $q'$, is positive (negative)
if the corresponding operator on the l.h.s. annihilates (creates)
a particle, the rule for the sign in the second bracket is the opposite
(compare previous Eqs. (\ref{DFSalpha}), (\ref{DFSbetaplus}), and
(\ref{DFSbetaminus})). The actual calculation of $W_{q'q}(\omega)$
with the help of the resolvent is described in Appendix \ref{DFSappCalcWq}.

The general expressions for the Green's functions still look like
before (see Eqs. (\ref{DFSctc}), (\ref{DFSpsigr}), (\ref{DFSexpphi}),
(\ref{DFSphicorr}) for the single-particle Green's function). However,
now the evaluation of the correlator (\ref{DFSphicorr}) of $\ph$
will yield contributions for all combinations of $q,q'$:

\begin{eqnarray}
\left\langle \ph(x',t)\ph(x,0)\right\rangle \nonumber \\
=-\sum_{q,q'=1}^{\infty}\frac{1}{\sqrt{q'q}}\left\langle (e^{iq'x'}\hat{b}_{q'}(t)-h.c.)(e^{iqx}\hat{b}_{q}-h.c.)\right\rangle \,.\label{DFSNEWphicorr}\end{eqnarray}
Together with the results for the $\hat{b}_{q}$-correlators, Eq.
(\ref{DFSnewbb}), this can be used to evaluate the Green's function,
which would work exactly as before in principle (expanding the exponential
in terms of $\exp(ix),\,\exp(ix')$). However, now one would have
to deal with far more terms. We have not carried through this calculation
in the general case so far.

Nevertheless, interesting behaviour of the fermions is already found
in the weak-coupling limit. In this limit, we neglect the effective
coupling between boson modes that has been induced by the bath (see
Appendix \ref{DFSappCalcWq}) and describe the correlator of each
boson mode separately as a damped oscillation, with:

\begin{equation}
\left\langle \hat{b}_{q}(t)\hat{b}_{q}^{\dagger}\right\rangle \approx e^{-i(q\omega_{0}+Re\,\delta\Omega_{-q})t-Im\,\delta\Omega_{-q}|t|}\,.\label{DFSnewbb}\end{equation}
Here the complex-valued frequency shift $\delta\Omega_{-q}$ is given
by Eq. (\ref{DFSappqqComplexFreqShift}) of Appendix \ref{DFSappCalcWq}.
In particular, the decay rate is

\begin{equation}
\textrm{Im}\,\delta\Omega_{-q}=qf_{q}^{2}\frac{N}{m\omega_{0}}\frac{\pi}{2}\left\langle \hat{F}\hat{F}\right\rangle _{q\omega_{0}}\,.\end{equation}
As expected, this is equal to one-half the decay rate $\Gamma_{q}$
given in Eq. (\ref{DFSdecayrateq}), which describes decay of populations
(rather than Green's functions).

For the special case of the Ohmic bath spectrum, Eq. (\ref{DFSohmicbath}),
(and with constant $f_{q}=1$), we obtain a decay rate rising quadratically
with excitation $q$,

\begin{equation}
\textrm{Im}\,\delta\Omega_{-q}=q^{2}\frac{N\gamma}{2}\,,\end{equation}
while the frequency shift is linear in $q$, up to corrections on
the order of $(q\omega_{0}/\omega_{c})^{2}$:

\begin{eqnarray}
Re\,\delta\Omega_{-q}=\frac{qN\gamma}{\pi\omega_{0}}\left\{ -\omega_{c}+\frac{q\omega_{0}}{2}\ln\frac{1+(q\omega_{0})/\omega_{c}}{1-(q\omega_{0})/\omega_{c}}\right\} \nonumber \\
\approx-q\frac{\omega_{c}}{\omega_{0}}\frac{N\gamma}{\pi}\,.\end{eqnarray}
These expressions hold for $q\omega_{0}$ in the {}``scaling'' region,
where the bath spectrum rises linearly with frequency (in our case,
for the sharp cutoff, $q\omega_{0}<\omega_{c}$). The last approximation
requires $q\omega_{0}\ll\omega_{c}$. 

The linear rise of the frequency shift with $q$ will prove convenient
in the following calculation, since it implies that the original frequency
$\omega_{0}$ is just shifted by $-(\omega_{c}/\omega_{0})N\gamma/\pi$,
for all boson modes with sufficiently small $q$. Concerning the decay
rate, the quadratic increase is expected, since in a naive Golden
Rule picture the decay of a particle from state $n_{F}+\delta n+1$
should be produced by all the transitions from $q=1$ up to $q=\delta n$.
Adding up their rates (which grow like $\left\langle \hat{F}\hat{F}\right\rangle _{q\omega_{0}}\propto q$
in the case of the Ohmic bath) leads to a total rate $\propto\delta n^{2}$,
which corresponds to the decay rate of the highest boson mode that
is excited by adding this particle. 

We will now confirm this qualitative picture by analyzing the weak-coupling
form of the single-particle Green's function for the Ohmic bath, at
$T=0$ (with $f_{q}\equiv1$ up to some cutoff). We must assume the
weak-coupling result to be a good approximation for all $q$. The
approximations introduced above yield:

\begin{equation}
\left\langle \ph(x',t)\ph(x,0)\right\rangle \approx\sum_{q=1}^{\infty}\frac{1}{q}e^{iq(x'-x)}e^{-i\omega'_{0}qt}e^{-N\gamma q^{2}|t|/2}\,.\label{DFSphiNewApprox}\end{equation}

Here the shifted frequency is given by $\omega'_{0}\equiv\omega_{0}-(\omega_{c}/\omega_{0})N\gamma/\pi$.
Note that, according to the discussion above, the simple form of the
correlator used here will be valid only for sufficiently small $q$,
where the conditions $q\omega_{0}\ll\omega_{c}$, $f_{q}=1$ hold.
In spite of this, we have extended the sum in (\ref{DFSphiNewApprox})
over all $q$. We have been able to do so because only $q$-values
up to the excitation level $\delta n$ will enter the end-result to
be derived below. The condition about $q$ being small is therefore
a condition on the excitation considered. 

\begin{figure}
\begin{center}\includegraphics[%
  width=0.90\columnwidth]{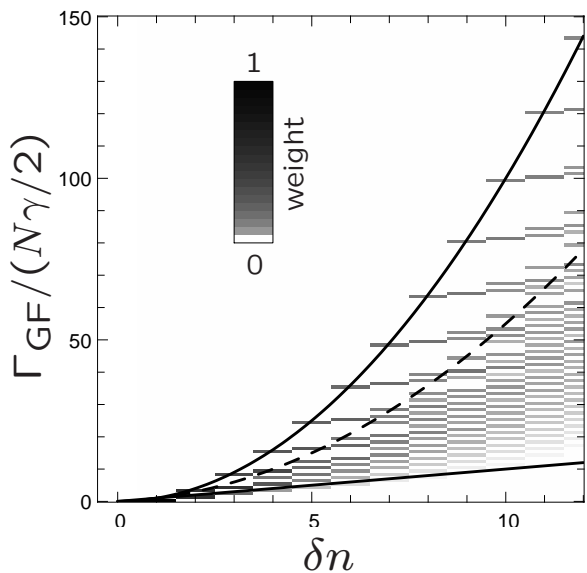}\end{center}

\caption{\label{DFSdistrDecayRates}Distribution of decay rates $\Gamma_{{\rm GF}}$
entering the Green's function in the weak-coupling approximation given
by Eq. (\ref{DFSnewGreensfunction}), for the Ohmic bath at $T=0$,
where the bath connects arbitrary oscillator levels. Decay rates are
plotted as a function of excitation level $\delta n$, and the grey
level visualizes the weight in Eq. (\ref{DFSnewGreensfunction}).
The full lines represent lower and upper bounds $N\gamma\delta n/2$
and $N\gamma\delta n^{2}/2$, while the dashed line is $N\gamma\delta n(\delta n+1)/4$. }
\end{figure}

By inserting (\ref{DFSphiNewApprox}) into the exponential (\ref{DFSexpphi})
and expanding it in the usual way, we obtain for the particle propagator
in this approximation:

\begin{eqnarray}
\left\langle \hat{c}_{n_{F}+1+\delta n}(t)\hat{c}_{n_{F}+1+\delta n}^{\dagger}\right\rangle =e^{-i\omega_{0}(n_{F}+1)t}\times\nonumber \\
e^{-i\omega'_{0}\delta nt}\sum_{m=1}^{\delta n}\frac{1}{m!}\sum_{q_{1},..,q_{m}}\frac{1}{q_{1}\cdot\ldots\cdot q_{m}}e^{-N\gamma(q_{1}^{2}+\ldots+q_{m}^{2})|t|/2}\label{DFSnewGreensfunction}\end{eqnarray}
Here the sum over $q_{1},\ldots,q_{m}$ is restricted by $\sum_{j=1}^{m}q_{j}=\delta n$
and $q_{j}\geq1$ . Therefore, we obtain contributions from decay
rates between $(N\gamma/2)\delta n$ and $(N\gamma/2)\delta n^{2}$.
The naive calculation (applying Golden Rule to Eq. (\ref{DFSnewcoupling})
and adding up all the transitions that are not blocked by the Pauli
principle) would give the following decay rate for the population
of level $n_{F}+1+\delta n$:

\begin{equation}
2\pi\frac{N}{2m\omega_{0}}\sum_{q=1}^{\delta n}\left\langle \hat{F}\hat{F}\right\rangle _{q\omega_{0}}^{T=0}=N\gamma\,\delta n(\delta n+1)/2\,.\end{equation}
In order to compare with the decay rate of the Green's function, we
divide by $2$ and obtain $(N\gamma/2)\delta n(\delta n+1)/2$, which
is between the lower and upper bounds derived in Eq. (\ref{DFSnewGreensfunction}).
The distribution of decay rates, involving the proper weights from
(\ref{DFSnewGreensfunction}), is plotted in Fig. \ref{DFSdistrDecayRates}. 

\begin{figure}
\begin{center}\includegraphics[%
  width=0.90\columnwidth]{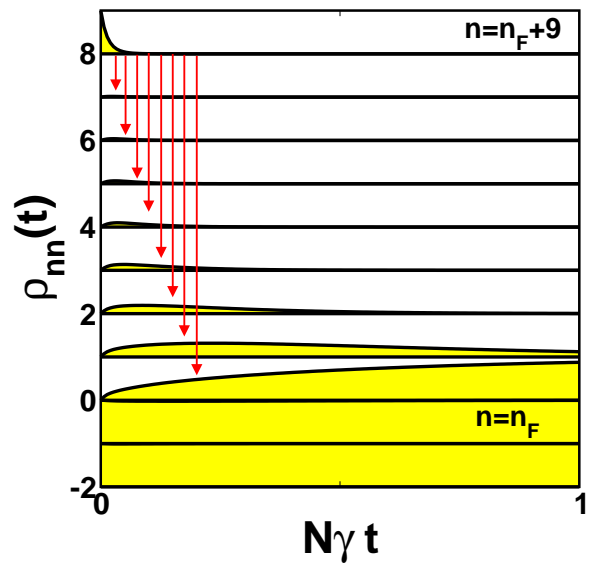}\end{center}

\caption{\label{DFSfigpopalldecay}Decay of populations for a bath that induces
transitions between levels that are arbitrarily far apart (indicated
by arrows); calculated in the weak-coupling approximation explained
in the text, for the Ohmic bath at $T=0$ (see Section \ref{DFSrelaxarbitraryenergytransfer}
and Appendix \ref{DFSappEvalTwoPGF}). }
\end{figure}
The same approximation can be used to obtain a weak-coupling version
of the two-particle Green's function (and, thus, the time-evolution
of the reduced single-particle density matrix, compare Sec. \ref{DFStwoparticlemaintext}).
The detailed steps of the evaluation are explained in Appendix \ref{DFSappEvalTwoPGF}.
An example of the resulting time-evolution of the populations is shown
in Fig. \ref{DFSfigpopalldecay}: The relaxation towards the $N+1$-particle
ground state is complete, the system is ergodic. At intermediate times,
heating of the Fermi surface is also observed, although it is much
less pronounced than in the example of Fig. \ref{DFSpopulationsdecay}.

\section{Conclusions}

We have analyzed a model system of fermions in a harmonic oscillator,
that are subject to a fluctuating force deriving from a linear bath
of oscillators. The force couples to the center-of-mass motion of
the fermions. This is a generalization of the Caldeira-Leggett model
of a single particle in a damped harmonic oscillator, and it is, in
principle, exactly solvable. However, here we have analyzed the limit
of large particle numbers, using the method of bosonization, i.e.
we have considered particle-hole excitations around the Fermi surface.
One of the boson modes corresponds to the center-of-mass motion, experiencing
dephasing and damping, while the others belong to the relative motion
which forms a kind of {}``decoherence-free subspace''. We have derived
exact (for $N\rightarrow\infty$) analytic expressions for Green's
functions, that can be obtained from the coefficients of a power series
expansion. These have been evaluated with the help of a symbolic computer
algebra program. The single-particle Green's function has been discussed
in detail, showing the smearing of the Fermi surface, the effect of
Pauli blocking, the dependence of level shapes on the bath spectrum
and the coupling strength, as well as the appearance of coherent peaks,
due to the relative motion. Based on the expression for the two-particle
Green's function, we have analyzed the decay of an excited state created
by adding one particle above the Fermi surface, where one can observe
the {}``heating'' around the Fermi surface (due to the effective
interaction between particles), as well as the incomplete decay of
the excited particle. We have provided simplified expressions for
the limiting case of high initial excitation energy. In addition,
we have discussed the time-evolution of the single-particle density
matrix for an extra particle that has been added in a coherent superposition
of excited states. Finally, we have extended our analysis to a more
general coupling, where the bath induces transitions between oscillator
levels that are not adjacent. In that case, the influence of the Pauli
principle becomes even more pronounced.

\begin{acknowledgments}
We thank C. Bruder, W. Belzig, H. Grabert, D. Loss and A. Zaikin for
comments and discussions. The work of F.M. has been supported by the
Swiss NSF and a DFG grant.
\end{acknowledgments}
\appendix

\section{Coupled oscillator problem}

\label{DFSappendixDiagon}We define the annihilation operators of
the bath modes (for $j\geq1$) as

\begin{equation}
\hat{d}_{j}=\sqrt{\frac{m\Omega_{j}}{2}}(\hat{Q}_{j}+i\frac{\hat{P}_{j}}{m\Omega_{j}})\,\Rightarrow\,\hat{Q}_{j}=(\hat{d}_{j}+\hat{d}_{j}^{\dagger})/\sqrt{2m\Omega_{j}}\,,\label{DFSdop}\end{equation}
and $\hat{d}_{0}\equiv\hat{b}_{1}$, which makes $\hat{Q}_{0}$ and
$\hat{P}_{0}$ the canonical variables of the c.m. mode and the coupling
in Eq. (\ref{DFShapprox}) equal to $\sqrt{N}\hat{F}\hat{Q}_{0}$. 

We now have to diagonalize the following quadratic Hamiltonian, according
to Eqs. (\ref{DFShapprox}) and (\ref{DFSfft0}):

\begin{equation}
\hat{H}'=\frac{1}{2m}P^{t}P+\frac{m}{2}Q^{t}(\Omega^{2}+V)Q\,.\label{DFShprime}\end{equation}
Here $P$ and $Q$ are (column) vectors, with operator entries $\hat{P}_{j}$,
$\hat{Q}_{j}$ ($j=0\ldots N_{B}$), where $j>0$ refers to the bath
oscillators, and $j=0$ refers to the $q=1$ mode. The diagonal matrix
$\Omega^{2}$ contains the eigenvalues $\Omega_{j}^{2}$ of the original
problem (with $\Omega_{0}\equiv\omega_{0}$) and $V$ is the coupling
matrix, with the nonvanishing entries $V_{j0}=V_{0j}=(g/m)\sqrt{N/N_{B}}$
($j>0$). 

We denote by $C$ the orthogonal matrix that diagonalizes the problem
of coupled oscillators defined in (\ref{DFShprime}), such that

\begin{equation}
C^{t}C=1,\, C^{t}(\Omega^{2}+V)C=\tilde{\Omega}^{2}\,,\label{DFSdiagonC}\end{equation}
with a diagonal matrix $\tilde{\Omega}^{2}$ containing the new eigenfrequencies.
Then we can express the old coordinates, momenta and boson operators
in terms of the new normal modes:

\begin{eqnarray}
Q & = & C\tilde{Q}\label{DFSQ}\\
P & = & C\tilde{P}\label{DFSP}\\
\tilde{d} & = & \sqrt{\frac{m\tilde{\Omega}}{2}}(\tilde{Q}+i\frac{\tilde{P}}{m\tilde{\Omega}})\,,\label{DFSdtildedef}\end{eqnarray}
where $\tilde{d}$ denotes a column vector whose entries are the operators
$\tilde{\hat{d}}_{j}$. In order to evaluate the correlator of the
c.m. mode, we have to relate the old annihilation operators $\hat{d}_{j}$
to the new ones. This is accomplished by inserting the relations (\ref{DFSQ}),
(\ref{DFSP}) into (\ref{DFSdop}) and expressing $\tilde{Q},\,\tilde{P}$
by $\tilde{d}$ according to (\ref{DFSdtildedef}). In matrix notation,
the result reads:

\begin{eqnarray}
d=\sqrt{\frac{m\Omega}{2}}(Q+i\frac{P}{m\Omega})=\nonumber \\
\frac{1}{2}\left(\sqrt{\Omega}C\frac{1}{\sqrt{\tilde{\Omega}}}-\frac{1}{\sqrt{\Omega}}C\sqrt{\tilde{\Omega}}\right)\tilde{d}^{\dagger}+\nonumber \\
\frac{1}{2}\left(\sqrt{\Omega}C\frac{1}{\sqrt{\tilde{\Omega}}}+\frac{1}{\sqrt{\Omega}}C\sqrt{\tilde{\Omega}}\right)\tilde{d}\,.\label{DFSdnew}\end{eqnarray}
Here $\tilde{d}^{\dagger}$ is to be understood as the column-vector
containing the hermitian conjugate operators as entries. Correlators
of the form $\left\langle \hat{b}_{1}(t)\hat{b}_{1}^{\dagger}\right\rangle $
are then obtained by inserting the result for $\hat{d}_{0}=\hat{b}_{1}$
and using the time-evolution of the {}``good'' boson operators $\tilde{\hat{d}}(t)=e^{-i\tilde{\Omega}t}\tilde{\hat{d}}$,
as well as their thermal equilibrium expectation values:

\begin{eqnarray}
\left\langle \hat{b}_{1}(t)\hat{b}_{1}\right\rangle =\left\langle \hat{d}_{0}(t)\hat{d}_{0}\right\rangle =\nonumber \\
\frac{1}{4}\left\langle \left[\left(\sqrt{\Omega}C\frac{1}{\sqrt{\tilde{\Omega}}}+\frac{1}{\sqrt{\Omega}}C\sqrt{\tilde{\Omega}}\right)\tilde{d}(t)\right]_{0}\right.\times\nonumber \\
\left.\left[\left(\sqrt{\Omega}C\frac{1}{\sqrt{\tilde{\Omega}}}-\frac{1}{\sqrt{\Omega}}C\sqrt{\tilde{\Omega}}\right)\tilde{d}^{\dagger}\right]_{0}\right\rangle  & =\nonumber \\
\frac{1}{4}\sum_{j=0}^{N_{B}}C_{0j}^{2}\left(\frac{\omega_{0}}{\tilde{\Omega}_{j}}-\frac{\tilde{\Omega}_{j}}{\omega_{0}}\right)e^{-i\tilde{\Omega}_{j}t}\,.\label{APPgrfct}\end{eqnarray}
At $T>0$ we would have to consider $\left\langle \tilde{\hat{d}}_{j}^{\dagger}(t)\tilde{\hat{d}}_{j}\right\rangle \neq0$
as well. 

We now turn to the actual diagonalization of the problem defined in
Eq. (\ref{DFShprime}), i.e. of $\Omega^{2}+V$ with $V_{0j}=V_{j0}=(g/m)\sqrt{N/N_{B}}$
for $j>0$ and all other entries of $V$ equal to zero. In fact, for
our purposes we only need the spectral weight of the mode $j=0$ of
the original problem in terms of the new eigenmodes and eigenfrequencies:

\begin{equation}
W(\omega)=\sum_{j=0}^{N_{B}}C_{0j}^{2}\delta(\omega-\tilde{\Omega}_{j})\,.\end{equation}
This is related directly to the resolvent

\begin{equation}
R(\epsilon)=\left[\frac{1}{\epsilon-(\Omega^{2}+V)}\right]_{00}\,,\end{equation}
since

\begin{equation}
R(\epsilon-i0)-R(\epsilon+i0)=2\pi i\sum_{j=0}^{N_{B}}C_{0j}^{2}\delta(\epsilon-\tilde{\Omega}_{j}^{2})\,,\end{equation}
where $\tilde{\Omega}_{j}^{2}$ is the $j$-th eigenvalue of the matrix
$\Omega^{2}+V$. Therefore, we have

\begin{equation}
W(\omega)=\frac{\omega}{\pi i}(R(\omega^{2}-i0)-R(\omega^{2}+i0))\theta(\omega)\,,\label{APPDFSWdef}\end{equation}
where we used $\delta(\omega-\tilde{\Omega}_{j})=2\omega\theta(\omega)\delta(\omega^{2}-\tilde{\Omega}_{j}^{2})$.

The evaluation of the resolvent $R(\epsilon)$ is straightforward,
because the {}``perturbation'' $V$ only connects $0$ and $j>0$.
Using $G^{-1}=\epsilon-\Omega^{2}$, we get

\begin{eqnarray}
R(\epsilon)=\left[G+GVG+GVGVG+\ldots\right]_{00}\nonumber \\
=G_{00}\sum_{n=0}^{\infty}\left(\sum_{j=1}^{N_{B}}V_{j0}^{2}G_{jj}G_{00}\right)^{n}\nonumber \\
=\left[\epsilon-\Omega_{0}^{2}-\sum_{j>0}\frac{V_{j0}^{2}}{\epsilon-\Omega_{j}^{2}}\right]^{-1}.\label{DFSresolvCalcOnew}\end{eqnarray}
In the continuum limit, we may use Eq. (\ref{DFSfft0}) to evaluate

\begin{eqnarray}
\sum_{j>0}\frac{V_{j0}^{2}}{\epsilon+i\delta-\Omega_{j}^{2}}=2\frac{N}{m}\int_{0}^{\infty}\frac{\left\langle \hat{F}\hat{F}\right\rangle _{\omega}^{T=0}\omega\, d\omega}{\epsilon+i\delta-\omega^{2}}\nonumber \\
\equiv\Delta(\epsilon)-\frac{i}{2}\Gamma(\epsilon)\,{\rm sgn}(\delta)\,.\label{APPDFSsumV}\end{eqnarray}
Here the definitions

\begin{eqnarray}
\Gamma(\epsilon) & = & 2\pi\frac{N}{m}\left\langle \hat{F}\hat{F}\right\rangle _{\sqrt{\epsilon}}^{T=0}\theta(\epsilon)\label{DFSappGammaDef}\\
\Delta(\epsilon) & = & \frac{1}{2\pi}\int\frac{\Gamma(\nu)}{\epsilon-\nu}d\nu\,.\end{eqnarray}
have been introduced. (where a principal value integral is implied
in the last line). Using this, we finally obtain (with (\ref{APPDFSWdef})-(\ref{APPDFSsumV})):

\begin{equation}
W(\omega)=\frac{\omega}{\pi}\theta(\omega)\frac{\Gamma(\omega^{2})}{(\omega^{2}-\omega_{0}^{2}-\Delta(\omega^{2}))^{2}+(\Gamma(\omega^{2})/2)^{2}}\,.\label{DFSWresult}\end{equation}

\section{Two-particle Green's function}

\label{DFStwoparticleappendix}The fermion operators $\hat{c}_{n}$
in the oscillator are related to the auxiliary fermion operators $\hat{\psi}(x)$
via Eq. (\ref{DFScpsi}). Therefore, the special two-particle Green's
function we need is given by: 

\begin{eqnarray}
\left\langle \hat{c}_{n}(0)\hat{c}_{l'}^{\dagger}(t)\hat{c}_{l}(t)\hat{c}_{n'}^{\dagger}(0)\right\rangle =\nonumber \\
\frac{1}{(2\pi)^{2}}\int dxdx'dydy'\, e^{i(l'x'+n'y'-lx-ny)}\times\nonumber \\
\left\langle \hat{\psi}(y,0)\hat{\psi}^{\dagger}(x',t)\hat{\psi}(x,t)\hat{\psi}^{\dagger}(y',0)\right\rangle \label{DFS2pgfviapsi}\end{eqnarray}
We use the expression of $\hat{\psi}$ in terms of the boson field
$\hat{\phi}$, Eq. (\ref{DFSpsibos}), to obtain (compare the single-particle
expressions Eqs. (\ref{DFSpsigr}), (\ref{DFSexpphi})):

\begin{eqnarray}
\left\langle \hat{\psi}(y,0)\hat{\psi}^{\dagger}(x',t)\hat{\psi}(x,t)\hat{\psi}^{\dagger}(y',0)\right\rangle =\nonumber \\
\frac{1}{(2\pi)^{2}}e^{i(n_{F}+1)(x+y-x'-y')}e^{E_{(2)}}\,,\label{DFSpsiexpE}\end{eqnarray}
with the exponent

\begin{eqnarray}
E_{(2)}\equiv\left\langle \hat{\phi}(y,0)\ph(x',t)\right\rangle -\left\langle \hat{\phi}(y,0)\ph(x,t)\right\rangle +\nonumber \\
\left\langle \hat{\phi}(y,0)\ph(y',0)\right\rangle +\left\langle \hat{\phi}(x',t)\ph(x,t)\right\rangle -\nonumber \\
\left\langle \hat{\phi}(x',t)\ph(y',0)\right\rangle +\left\langle \hat{\phi}(x,t)\ph(y',0)\right\rangle -\nonumber \\
\frac{1}{2}(C_{x}+C_{y}+C_{x'}+C_{y'})\,.\label{DFSfirstexpon}\end{eqnarray}
Here we have introduced $C_{x}\equiv\left\langle \hat{\phi}(x,0)^{2}\right\rangle -\left\langle \hat{\phi}(x,0)^{2}\right\rangle _{(0)}$,
where the second correlator refers to the non-interacting case (without
bath) and stems from the factor $r$ (see Eq. (\ref{DFSpsibos})).

Once again, we restrict the further evaluation to the special case
of $T=0$, where some of the expressions become slightly simpler.
We express the nontrivial c.m. mode contribution to the correlator
of $\ph$ in terms of the $\hat{b}_{1}^{(\dagger)}$-correlators $\alpha(t),\,\beta_{\pm}(t)$
introduced above (see Eqs. (\ref{DFSbetaplus}), (\ref{DFSbetaminus})
and (\ref{DFSabb})). Since the two-particle Green's function involving
the oscillator fermion operators $\hat{c}_{n}$ is related to that
of the auxiliary fermion operators $\hat{\psi}$ by means of a Fourier
integral over the coordinates $x,y,x',y'$ (cf. Eq. (\ref{DFS2pgfviapsi})),
we find it convenient to introduce the following abbreviations (similar
to our approach for the single-particle Green's function):

\begin{equation}
X^{(')}\equiv e^{ix^{(')}},\, Y^{(')}\equiv e^{iy^{(')}}\,\end{equation}
Then we obtain

\begin{equation}
C_{x}=\beta_{+}(0)+\beta_{-}(0)-1-\alpha(0)(X^{2}+X^{-2})\,,\end{equation}
as well as the form of the $\ph$-correlator given already in Eq.
(\ref{DFSphiXX}).

Therefore, the exponent $E_{(2)}$ (\ref{DFSfirstexpon}) is given
as a sum of terms containing $\alpha,\beta_{\pm}$ and binomials constructed
out of $X,X',Y$ and $Y'$, as well as non-interacting contributions
$\left\langle \ph(x',t)\ph(x,0)\right\rangle _{(0)}$, that also contain
higher powers of these variables (see Eq. (\ref{DFSphi0qsum})). According
to Eq. (\ref{DFS2pgfviapsi}), the two-particle Green's function

\begin{equation}
\left\langle \hat{c}_{n}(0)\hat{c}_{l'}^{\dagger}(t)\hat{c}_{l}(t)\hat{c}_{n'}^{\dagger}(0)\right\rangle \end{equation}
is obtained by expanding $\exp(E_{(2)})$ in a power series with respect
to $X,\, X',\, Y$ and $Y'$, and reading off the coefficient in front
of the component $X^{\delta l}X'^{-\delta l'}Y^{\delta n}Y'^{-\delta n'}$,
where $\delta l\equiv l-(n_{F}+1)$ etc. Using (\ref{DFS2pgfviapsi}),
(\ref{DFSpsiexpE}) and (\ref{DFSfirstexpon}), as well as splitting
off non-interacting terms, we find that the expression to be expanded
can be written in the following form:

\begin{equation}
A_{0}e^{\tilde{E}}\,.\label{DFSa0etilde}\end{equation}
Here $A_{0}$ yields the correct result for the non-interacting case
(cf. the application of Wick's theorem, Eq. (\ref{DFSwick})),

\begin{eqnarray}
A_{0}=\sum_{\tilde{l}<0}\left(\frac{X}{X'}\right)^{\tilde{l}}\sum_{\tilde{n}\geq0}\left(\frac{Y}{Y'}\right)^{\tilde{n}}+\nonumber \\
\sum_{\tilde{n}'\geq0}\left(\frac{X}{Y'}\right)^{\tilde{n}'}\sum_{\tilde{n}\geq0}\left(\frac{Y}{X'}\right)^{\tilde{n}}e^{i\omega_{0}(\tilde{n}-\tilde{n}')t}\,,\label{a0}\end{eqnarray}
and the exponent $\tilde{E}$ is a rather lengthy expression involving
$\alpha(t),\,\alpha(-t)=\alpha^{*}(t),\,\beta_{\pm}(t),\,\beta_{\pm}(-t)=\beta_{\pm}^{*}(t)$
and binomials made out of $X,X',Y$ and $Y'$:

\begin{eqnarray}
\tilde{E}\equiv\alpha(t)\,(X'Y'+\frac{1}{X'Y'}-XY'-\frac{1}{XY'})+\nonumber \\
\alpha^{*}(t)\,(XY+\frac{1}{XY}-X'Y-\frac{1}{X'Y})-\nonumber \\
\alpha(0)\,(XX'+\frac{1}{XX'}+YY'+\frac{1}{YY'})+\nonumber \\
\tilde{\beta}_{+}(t)\,(\frac{X}{Y'}-\frac{X'}{Y'})+\tilde{\beta}_{+}^{*}(t)\,(\frac{Y}{X'}-\frac{Y}{X})+\nonumber \\
\tilde{\beta}_{+}(0)\,(\frac{Y}{Y'}+\frac{X'}{X})+\nonumber \\
\beta_{-}(t)\,(\frac{Y'}{X}-\frac{Y'}{X'})+\beta_{-}^{*}(t)\,(\frac{X'}{Y}-\frac{X}{Y})+\nonumber \\
\beta_{-}(0)\,(\frac{Y'}{Y}+\frac{X}{X'})-\nonumber \\
\frac{1}{2}(C_{x}+C_{y}+C_{x'}+C_{y'})\,.\label{DFSetilde}\end{eqnarray}
In this equation, we have employed the abbreviation $\tilde{\beta}_{+}(t)\equiv\beta_{+}(t)-e^{-i\omega_{0}t}$.
Note that $\tilde{E}$ vanishes if the coupling to the bath is switched
off.

\section{Limiting expressions for two-particle Green's function}

\label{sec:Limiting-expressions-for}Here we will show how to derive
simplified expressions for the two-particle Green's function of the
previous section, in the limit of weak coupling and for high initial
excitation. We rewrite Eq. (\ref{DFSetilde}) for the weak coupling
case, setting $\alpha(t)=\beta_{-}(t)=0$, $\beta_{+}(0)=1$ and keeping
only the (slowly decaying) $\beta_{+}(t)$ (compare the main text,
Sec. \ref{DFStwoparticlemaintext}, and the corresponding discussion
for the single-particle case in Sec. \ref{DFSgreensfctT0}):

\begin{equation}
\tilde{E}=\nu(t)\left(\frac{X}{Y'}-\frac{X'}{Y'}\right)+\nu^{*}(t)\left(\frac{Y}{X'}-\frac{Y}{X}\right)\,.\label{EtildeWC}\end{equation}
Here we have set $\omega_{0}=0$, defining $\nu(t)=\tilde{\beta}_{+}(t)e^{i\omega_{0}t}=e^{-i\delta\omega t-N\gamma t/2}-1$.
We will also set $\omega_{0}=0$ inside $A_{0}$, because it can be
checked that $\omega_{0}$ enters the final density matrix only in
the trivial form $\exp(i\omega_{0}(n-n')t)$. In the following, we
will shift indices, denoting $\delta l\equiv l-(n_{F}+1)$ by $l$,
for brevity. Our task then is to obtain the coefficient of $X^{l}X'^{-l'}Y^{n}Y'^{-n'}$
in the expansion of $A_{0}\exp\tilde{E}$. 

Let us turn first to the product of the second part of $A_{0}$ (see
Eq. (\ref{a0})) with the exponential $\exp(\tilde{E})$. As we will
see, this describes the decay of the excited state, while the first
part will be connected to heating around the Fermi surface. Performing
the expansion leads to a coefficient

\begin{equation}
\sum\frac{\nu^{m_{1}+m_{2}}\nu^{*\tilde{m}_{1}+\tilde{m}_{2}}}{m_{1}!m_{2}!\tilde{m}_{1}!\tilde{m}_{2}!}(-1)^{m_{2}+\tilde{m}_{2}}\,,\label{C2expr}\end{equation}
where the summation is over all $m_{1,2},\tilde{m}_{1,2},\tilde{n},\tilde{n}'\geq0$,
subject to the constraints:

\begin{eqnarray}
l & = & \tilde{n}'+m_{1}-\tilde{m}_{2}\\
l' & = & \tilde{n}+\tilde{m}_{1}-m_{2}\\
n & = & \tilde{n}+\tilde{m}_{1}+\tilde{m}_{2}\\
n' & = & \tilde{n}'+m_{1}+m_{2}\,.\end{eqnarray}
This requires $n-l'=n'-l=m=m_{2}+\tilde{m}_{2}\ge0$, i.e. the Green's
function vanishes unless $(l,l')=(n',n)-m(1,1)$ for some $m\ge0$,
corresponding to the total energy lost to the bath (in units of $\hbar\omega_{0}$).
If the given $l,l',n,n'$ obey this relation, then we are left with
only three equations for six summation indices. The equations $0\leq\tilde{n}=l'+m_{2}-\tilde{m}_{1}$
and $0\leq\tilde{n}'=l+\tilde{m}_{2}-m_{1}$ can be fulfilled if $\tilde{m}_{1}\leq l'+m_{2}$
and $m_{1}\leq l+\tilde{m}_{2}$, respectively. The third equation
is $m_{2}+\tilde{m}_{2}=m$. Thus, we have a sum running over $m_{2}=0\ldots m$,
$m_{1}=0\ldots l+\tilde{m}_{2}$ and $\tilde{m}_{1}=0\ldots l'+m_{2}$,
with $\tilde{m}_{2}=m-m_{2}$:

\begin{equation}
(-1)^{m}\sum\frac{\nu^{m_{1}+m_{2}}\nu^{*\tilde{m}_{1}+\tilde{m}_{2}}}{m_{1}!m_{2}!\tilde{m}_{1}!\tilde{m}_{2}!}\,.\end{equation}
This coefficient depends on $(m,l,l')$. However, in the limit of
a high initial excitation, $n,n',l,l'\gg1$, we can extend the sums
over $m_{1}$ and $\tilde{m}_{1}$ to infinity. This yields:

\begin{equation}
\rho_{{\rm decay}}(m,t)\approx\frac{(-1)^{m}}{m!}(\nu(t)+\nu^{*}(t))^{m}e^{\nu(t)+\nu^{*}(t)}\,.\label{WCexplicitSecondPart}\end{equation}
In this limit, the presently discussed part of the two-particle Green's
function only depends on the net number of quanta $m$ emitted into
the bath, and its time-evolution is described by the simple expression
(\ref{WCexplicitSecondPart}). For $t\rightarrow\infty$, we have
$\nu(t)\rightarrow-1$, and thus $\rho_{{\rm decay}}(m,t)\rightarrow\frac{2^{m}}{m!}e^{-2}$. 

The product of the first part of $A_{0}$with $\exp(\tilde{E})$ describes
how some excitation above the Fermi surface is able to influence the
states near the surface, by creating some {}``heating'' (via the
bath-induced interaction between particles). Again, expanding leads
to a coefficient of the same form as Eq. (\ref{C2expr}), but with
different constraints:

\begin{eqnarray}
l & = & \tilde{l}+m_{1}-\tilde{m}_{2}\\
l' & = & \tilde{l}+\tilde{m}_{1}-m_{2}\\
n & = & \tilde{n}+\tilde{m}_{1}+\tilde{m}_{2}\\
n' & = & \tilde{n}+m_{1}+m_{2}\,,\end{eqnarray}
where $\tilde{l}<0$ and $\tilde{n}\geq0$. We find $n-l'=n'-l=\tilde{n}-\tilde{l}+\tilde{m}_{2}+m_{2}>0$.
The conditions $\tilde{l}=l+\tilde{m}_{2}-m_{1}<0$ and $\tilde{n}=n-(\tilde{m}_{1}+\tilde{m}_{2})\ge0$
can be fulfilled if $m_{1}>\tilde{m}_{2}+l$ and $\tilde{m}_{1}+\tilde{m}_{2}\leq n$.
Thus, the following sum runs over all $\tilde{m}_{2}=0\ldots n$,
$\tilde{m}_{1}=\textrm{max}(0,l'+1)\ldots n-\tilde{m}_{2}$, $m_{1}=\textrm{max}(0,\tilde{m}_{2}+l+1)\ldots l-l'+\tilde{m}_{1}+\tilde{m}_{2}$,
with $m_{2}=\tilde{m}_{1}+\tilde{m}_{2}-m_{1}+l-l'$:

\begin{equation}
\rho_{{\rm heat}}=\nu^{l-l'}\sum\frac{(\nu\nu^{*})^{\tilde{m}_{1}+\tilde{m}_{2}}}{m_{1}!m_{2}!\tilde{m}_{1}!\tilde{m}_{2}!}\,(-1)^{m_{2}+\tilde{m}_{2}}.\end{equation}
In the limit $n,n'\rightarrow\infty$, the sums over $\tilde{m}_{2}$
and $\tilde{m}_{1}$ become unbounded. Unfortunately, no further simplification
is possible. Note that we are interested in small $l,l'$, describing
states near the Fermi surface, which are affected by the high excitation
at $(n,n')$. In this limit, the dependence on $n,n'$ drops out,
and we have $\rho_{{\rm heat}}=\rho_{{\rm heat}}(l,l',t)$.

\section{Resolvent for arbitrary coupling between levels}

\label{DFSappCalcWq}For the case of damping of all boson modes (transitions
with arbitrary energy transfer, see Sec. \ref{DFSrelaxarbitraryenergytransfer}),
we introduce the resolvent

\begin{equation}
R_{q'q}(\epsilon)=\left[\frac{1}{\epsilon-(\Omega^{2}+V)}\right]_{-q',-q}\,,\end{equation}
in terms of which the {}``spectral weight'' $W_{q'q}(\omega)$ reads
(for $\omega\geq0$)

\begin{equation}
W_{q'q}(\omega)=\frac{2\omega}{\pi}\textrm{Im}\, R_{q'q}(\omega^{2}-i0^{+})\,.\label{DFSappqqWexpr}\end{equation}
According to Eq. (\ref{DFSnewV}), the perturbation in the problem
of coupled oscillators is given by $V_{-q,j}=V_{j,-q}=vf_{q}q$ ($q\geq1,\, j\geq1,\, v\equiv g\sqrt{N/N_{B}}/m$),
i.e. it always leads to transitions between bath modes and boson modes
(note there are no nonvanishing terms $V_{j,j}$ or $V_{-q,-q'}$).
This allows us to employ a procedure similar to the previous one (Eq.
(\ref{DFSresolvCalcOnew})), and to sum the geometric series, in order
to obtain the following exact result for the resolvent:

\begin{equation}
R_{q'q}(\epsilon)=\delta_{q',q}G_{-q}+\frac{G_{-q'}G_{-q}q'qf_{q'}f_{q}}{\left(v^{2}\sum_{j\geq1}G_{j}\right)^{-1}-\sum_{\tilde{q}\geq1}\tilde{q}^{2}f_{\tilde{q}}^{2}G_{-\tilde{q}}}\,.\end{equation}
Here $G_{k}\equiv G_{kk}=(\epsilon-\Omega_{k}^{2})^{-1}$ is the unperturbed
resolvent. The sum over bath modes $j\geq1$ may be expressed in terms
of the bath spectrum (see Eq. (\ref{DFSfft0}) in main text). This
yields, finally:

\begin{eqnarray}
R_{q'q}(\epsilon)=\frac{\delta_{q',q}}{\epsilon-(q\omega_{0})^{2}}+\frac{f_{q'}q'}{\epsilon-(q'\omega_{0})^{2}}\frac{f_{q}q}{\epsilon-(q\omega_{0})^{2}}\times\nonumber \\
\left[\lambda^{-1}-\sum_{\tilde{q}\geq1}\frac{f_{\tilde{q}}^{2}\tilde{q}^{2}}{\epsilon-(\tilde{q}\omega_{0})^{2}}\right]^{-1}\,.\label{DFSappnewResolvent}\end{eqnarray}
We have defined:

\begin{equation}
\lambda\equiv\frac{2N}{m}\int_{0}^{\infty}d\omega\,\frac{\omega\left\langle \hat{F}\hat{F}\right\rangle _{\omega}^{T=0}}{\epsilon-\omega^{2}}\,.\end{equation}
The steps up to this point correspond to integrating out the bath,
which leads to a problem of coupled, damped boson modes (whose spectrum
is given indirectly, via the poles of the resolvent). 

In order to analyze the weak-coupling limit of this expression, we
treat $\lambda$ as small. Then we obtain the following result for
the off-diagonal terms of the resolvent ($q'\neq q$), which describe
the induced coupling between the boson modes:

\begin{eqnarray}
 &  & R_{q'q}(\epsilon)\approx\lambda\, f_{q'}f_{q}q'q\frac{1}{\epsilon-(q'\omega_{0})^{2}-\lambda f_{q'}^{2}q'^{2}}\times\nonumber \\
 &  & \frac{1}{\epsilon-(q\omega_{0})^{2}-\lambda f_{q}^{2}q^{2}}\,.\end{eqnarray}
This expression has been derived by keeping only the poles near $q\omega_{0}$
and $q'\omega_{0}$ and neglecting terms of order $\lambda^{2}$ in
the denominator of Eq. (\ref{DFSappnewResolvent}). The prefactor
$\lambda$ makes these off-diagonal contributions small for small
system-bath coupling.

On the other hand, the diagonal contributions may be approximated
as:

\begin{equation}
R_{qq}(\epsilon)\approx\frac{1}{\epsilon-(q\omega_{0})^{2}-\lambda f_{q}^{2}q^{2}}\,.\label{DFSappqqRqq}\end{equation}
Here the approximation consisted in dropping poles other than that
at $q\omega_{0}$ in the sum over $\tilde{q}$ in expression (\ref{DFSappnewResolvent}).

Inserting this approximation (\ref{DFSappqqRqq}) into Eq. (\ref{DFSappqqWexpr})
for $W_{qq}$, we obtain:

\begin{eqnarray}
W_{qq}(\omega)\approx\frac{2\omega}{\pi}\,\textrm{Im}\,\left(\omega^{2}-i0^{+}-(q\omega_{0})^{2}-\lambda f_{q}^{2}q^{2}\right)^{-1}\nonumber \\
\approx\frac{1}{\pi}\frac{-\textrm{Im}\,\delta\Omega_{-q}}{(\omega-q\omega_{0}-\textrm{Re}\,\delta\Omega_{-q})^{2}+(\textrm{Im}\,\delta\Omega_{-q})^{2}}\,.\label{DFSappqqWapprox}\end{eqnarray}
Here, $\lambda$ in the first line is to be evaluated at $\epsilon=\omega^{2}-i0^{+}$.
In the second line, we have furthermore approximated $W_{qq}(\omega)$
as a Lorentz peak of width $\textrm{Im}\,\delta\Omega_{-q}$, where
the complex frequency shift $\delta\Omega_{-q}$ of boson mode $q$
is given as:

\begin{equation}
\delta\Omega_{-q}\equiv qf_{q}^{2}\frac{N}{m\omega_{0}}\int_{0}^{\infty}d\omega\,\frac{\omega\left\langle \hat{F}\hat{F}\right\rangle _{\omega}^{T=0}}{(q\omega_{0})^{2}-i0^{+}-\omega^{2}}\,.\label{DFSappqqComplexFreqShift}\end{equation}
The Fourier tansform of the weak-coupling result (\ref{DFSappqqWapprox})
for $W_{qq}$ is, therefore, an exponentially decaying oscillation,
$\exp[-i(q\omega_{0}+Re\,\delta\Omega_{-q})t-\textrm{Im}\,\delta\Omega_{-q}|t|]$.

\section{Evaluation of two-particle Green's function}

\label{DFSappEvalTwoPGF}In this appendix, we show the detailed steps
needed in the actual numerical evaluation of a two-particle Green's
function. We do this for the case of coupling between arbitrary levels
(see section \ref{DFSrelaxarbitraryenergytransfer}), for the Ohmic
bath at $T=0$, evaluated in the weak-coupling approximation given
by Eq. (\ref{DFSphiNewApprox}). The resulting approximation for the
two-particle Green's function can be obtained by inserting Eq. (\ref{DFSphiNewApprox})
of the main text into the expression for the exponent $E_{(2)}$ (Eq.
(\ref{DFSfirstexpon}) of Appendix \ref{DFStwoparticleappendix},
where $C_{x}=C_{x'}=C_{y}=C_{y'}=0$ in the weak-coupling case). In
writing down the result, we use the abbreviations $w\equiv\exp(-i\omega'_{0}t)$,
$u\equiv\exp(-\tilde{\gamma}|t|)$ (and $X\equiv\exp(ix)$ etc., as
before), as well as:

\begin{eqnarray}
A_{q} & \equiv & \frac{1}{q}\left(\frac{Y}{X'}\frac{1}{w}\right)^{q}u^{q^{2}}\label{DFSappADEF}\\
B_{q} & \equiv & -\frac{1}{q}\left(\frac{Y}{X}\frac{1}{w}\right)^{q}u^{q^{2}}\\
C_{q} & \equiv & -\frac{1}{q}\left(\frac{X'}{Y'}w\right)^{q}u^{q^{2}}\\
D_{q} & \equiv & \frac{1}{q}\left(\frac{X}{Y'}w\right)^{q}u^{q^{2}}\,.\label{DFSappDdef}\end{eqnarray}
Then we have:

\begin{eqnarray}
E_{(2)}\approx\sum_{q=1}^{\infty}A_{q}+B_{q}+C_{q}+D_{q}\nonumber \\
+\sum_{q=1}^{\infty}\frac{1}{q}\left\{ \left(\frac{Y}{Y'}\right)^{q}+\left(\frac{X'}{X}\right)^{q}\right\} \,.\label{DFSapproxE2}\end{eqnarray}
As explained in Appendix \ref{DFStwoparticleappendix}, the two-particle
Green's function

\begin{equation}
\left\langle \hat{c}_{n}(0)\hat{c}_{l'}^{\dagger}(t)\hat{c}_{l}(t)\hat{c}_{n'}^{\dagger}(0)\right\rangle \end{equation}
is given by the coefficient of the term $X^{\delta l}X'^{-\delta l'}Y^{\delta n}Y'^{-\delta n'}$
in the series generated from $\exp(E_{(2)})$, where $\delta l\equiv l-(n_{F}+1)$
etc. 

First, the last sum in (\ref{DFSapproxE2}) is split off from $E_{(2)}$
(as it does not depend on the time-dependent factors $w$ and $u$).
It yields:

\begin{eqnarray}
\exp\left[\sum_{q=1}^{\infty}\frac{1}{q}\left\{ \left(\frac{Y}{Y'}\right)^{q}+\left(\frac{X'}{X}\right)^{q}\right\} \right]=\nonumber \\
(1-\frac{Y}{Y'})^{-1}(1-\frac{X'}{X})^{-1}=\sum_{k_{x},k_{y}=0}^{\infty}\left(\frac{Y}{Y'}\right)^{k_{y}}\left(\frac{X'}{X}\right)^{k_{x}}\,.\label{DFSsplitoffexpon}\end{eqnarray}
We denote by $K(\tilde{l},\tilde{l}',\tilde{n},\tilde{n}')$ the coefficient
of $X^{\tilde{l}}X'^{-\tilde{l}'}Y^{\tilde{n}}Y'^{-\tilde{n}'}$ in
the expansion of the remaining exponential $\exp(\tilde{E}_{(2)})$
(where that sum has been left out in the exponent). According to Eq.
(\ref{DFSsplitoffexpon}), the desired coefficient of the \emph{full}
exponential $\exp(E_{(2)})$ is then given by summing the coefficients
$K$ over all $k_{x},\, k_{y}\geq0$ with

\begin{equation}
\tilde{l}=\delta l+k_{x},\,\tilde{l}'=\delta l'+k_{x},\,\tilde{n}=\delta n-k_{y},\,\tilde{n}'=\delta n'-k_{y}\,,\label{DFSoldrelations}\end{equation}
for given $\delta l,\,\delta l',\,\delta n,\,\delta n'$. 

Turning now to the evaluation of $K$ itself, we have

\begin{equation}
e^{\sum_{q=1}^{\infty}A_{q}}=\prod_{q=1}^{\infty}\left(\sum_{\alpha_{q}=0}^{\infty}\frac{A_{q}^{\alpha_{q}}}{\alpha_{q}!}\right)\,,\label{DFSappAexp}\end{equation}
and analogous expressions for the other three contributions to $\exp\tilde{E}_{(2)}$,
with corresponding sets of exponents $\beta_{q},\gamma_{q}$ and $\delta_{q}$.

We introduce $\alpha\equiv\sum_{q=1}^{\infty}q\alpha_{q}$, and analogous
definitions for $\beta,\gamma,\delta$. Then, by looking at Eqs. (\ref{DFSappAexp})
and the definitions (\ref{DFSappADEF})-(\ref{DFSappDdef}), we find
that the following relations have to be fulfilled in order to obtain
the desired coefficient $K$ of $X^{\tilde{l}}X'^{-\tilde{l}'}Y^{\tilde{n}}Y'^{-\tilde{n}'}$
in the expansion of $\exp(\tilde{E}_{(2)})$:

\begin{equation}
\tilde{l}=\delta-\beta,\,\tilde{l}'=\alpha-\gamma,\,\tilde{n}=\alpha+\beta,\,\tilde{n}'=\gamma+\delta\label{DFSrestrictionforExponents}\end{equation}
 Thus, the coefficient $K$ is calculated by adding up the following
contribution for each set of exponents $\alpha_{1},\alpha_{2},\ldots,\beta_{1},\ldots,\gamma_{1},\ldots,\delta_{1},\ldots\geq0$
that fulfills Eq. (\ref{DFSrestrictionforExponents}):

\begin{eqnarray}
 &  & \frac{(-1)^{\sum_{q}\beta_{q}+\gamma_{q}}}{\left(\prod_{q}q^{\alpha_{q}}\alpha_{q}!\right)\left(\prod_{q}q^{\beta_{q}}\beta_{q}!\right)\left(\prod_{q}q^{\gamma_{q}}\gamma_{q}!\right)\left(\prod_{q}q^{\delta_{q}}\delta_{q}!\right)}\times\nonumber \\
 &  & u^{\sum_{q}q^{2}(\alpha_{q}+\beta_{q}+\gamma_{q}+\delta_{q})}w^{\gamma+\delta-\alpha-\beta}.\label{DFSbigcoeff}\end{eqnarray}
According to Eq. (\ref{DFSoldrelations}), we have the upper bounds
$\tilde{n}=\alpha+\beta\leq\delta n$ and $\tilde{n}'=\gamma+\delta\leq\delta n'$
(and, consequently from Eq. (\ref{DFSrestrictionforExponents}), $|\tilde{l}|,|\tilde{l}'|\leq\max(\delta n,\delta n')$).
Together with $\alpha_{q},\beta_{q},\gamma_{q},\delta_{q}\geq0$ and
the definition of $\alpha$, this means we have to consider only $q$-values
up to $q=\delta n$ for $\alpha_{q},\,\beta_{q}$ and $q=\delta n'$
for $\gamma_{q},\delta_{q}$; all exponents for higher $q$ must vanish.
This justifies the use of the special Ohmic weak-coupling form (\ref{DFSsplitoffexpon})
of the exponent, as long as the initial excitation (described by $\delta n,\,\delta n'$)
is not too far from the Fermi level. For lower $q$, we have upper
bounds of $q\alpha_{q}\leq\delta n$ (and so on). 

In the actual numerical evaluation, the following procedure is used:
Given a maximum excitation of $\delta n_{{\rm max}}$, a table of
coefficients $K$ for all possible values of $\alpha,\,\beta,\,\gamma,\,\delta=0\ldots\delta n_{{\rm max}}$
is produced, by generating, for each set $(\alpha,\beta,\gamma,\delta)$
all possible $\alpha_{1},\alpha_{2},\ldots,\beta_{1},\ldots,\gamma_{1},\ldots,\delta_{1},\ldots$
that fulfill $\alpha\equiv\sum_{q=1}^{\infty}q\alpha_{q}$ etc. and
adding up the resulting contributions from (\ref{DFSbigcoeff}). Since
$w$ and $u$ are actually functions of time, the contributions are
not added up immediately but stored in a symbolic form, with the numerical
prefactor, the $u$-exponent and the $w$-exponent as entries. The
results may finally be used to calculate the two-particle Green's
function, for any given $\delta l,\delta l',\delta n,\delta n'\leq\delta n_{{\rm max}}$
and any given time $t$. It is given by: 

\begin{equation}
\sum_{k_{x},k_{y}\geq0}K(\tilde{l}=\delta l+k_{x},\,\tilde{l}'=\delta l'+k_{x},\,\tilde{n}=\delta n-k_{y},\,\tilde{n}'=\delta n'-k_{y})\,,\end{equation}
where $\alpha,\beta,\gamma,\delta$ are obtained from $\tilde{l}$
etc. by Eq. (\ref{DFSrestrictionforExponents}). Note that, because
of $\tilde{l},\tilde{l}'\leq\max(\delta n,\delta n')$ and $\tilde{n},\tilde{n}'\geq0$,
the summation variables $k_{x},k_{y}$ are also bounded from above.


\begin{thebibliography}{10}
\bibitem{footnoteWB}Our Eq. (\ref{GreensFunctExt}), at $t=0$, turns into Eq. (43) of
the first of Ref. \onlinecite{wonneberger}, if we set $\alpha_{1}^{{\rm WW}}=-\alpha(0)$,
$\gamma_{1}^{{\rm WW}}=\beta_{-}(0)$, $N^{{\rm WW}}=n_{F}$, and
$M^{{\rm WW}}=(n+n')/2,\, p^{{\rm WW}}=n'-n$, where $'{\rm WW}'$
refers to the notation used there. Note we used $1+0^{+}-e^{-ix}$
in the denominator to regularize, and do not need the explicit $1/2$
of Ref.\onlinecite{wonneberger}.
\end{thebibliography}
\end{document}